\documentclass[pre,twocolumn,showpacs,preprintnumbers,amsmath,amssymb,superscriptaddress]{revtex4}

\usepackage{mathrsfs}
\usepackage{amsmath}
\usepackage[colorlinks,linkcolor=blue,citecolor=blue]{hyperref}
\usepackage{multirow}

\begin{document}
\title{Jarzynski Equality, Crooks Fluctuation Theorem and the Fluctuation Theorems of Heat for Arbitrary Initial States}
\author{Zongping Gong}
\affiliation{School of Physics, Peking University, Beijing 100871, China}
\author{H. T. Quan}
\email{htquan@pku.edu.cn}
\affiliation{School of Physics, Peking University, Beijing 100871, China}
\affiliation{Collaborative Innovation Center of Quantum Matter, Beijing 100871, China}
\date{\today}
\begin{abstract}
By taking full advantage of the dynamic property imposed by the detailed balance condition, we derive a new refined unified fluctuation theorem (FT) for general stochastic thermodynamic systems. This FT involves the joint probability distribution functions of the final phase space point and a thermodynamic variable. 
Jarzynski equality, Crooks fluctuation theorem, and the FTs of heat as well as the trajectory entropy production can be regarded as special cases of this refined unified FT, and all of them 
are generalized to arbitrary initial distributions. 
We also find that the refined unified FT can easily reproduce the FTs for processes with the feedback control, due to its unconventional structure that separates the thermodynamic variable from the choices of initial distributions. Our result is heuristic for further understanding of the relations and distinctions between all kinds of FTs, and might be valuable for studying thermodynamic processes with information exchange.
\end{abstract}
 \pacs{05.70.Ln, 
  05.20.-y,	
  05.40.-a, 
  05.10.Gg	
}
\maketitle

\section{Introduction}
There have been many great progresses in nonequilibrium statistical physics of small systems in the past two decades. Compared with classical statistical physics where relative thermal fluctuations are generally Gaussian and vanishingly small, fluctuations become much more prominent in small systems that undergo processes arbitrarily far from equilibrium due to some nonequilibrium external drivings. Despite of the complexity that originates from the arbitrariness of the driving protocols, these fluctuations turn out to satisfy some strong, useful and elegant universal properties, exactly depicted by the fluctuation theorems (FTs) \cite{Jarzynski_1997,Jarzynski_1997b,Crooks_1999,Crooks_2000,Seifert_2005,Park_2014}. 
For example, the Jarzynski equality (JE) \cite{Jarzynski_1997}, which connects the work done in nonequilibrium processes and the free energy difference of the system at the initial and the final stages, is an integral fluctuation theorem (IFT) of work, while its stronger version, the Crooks fluctuation theorem (CFT) \cite{Crooks_1999}, is a detailed fluctuation theorem (DFT) of work. It should be emphasized that both the JE and the CFT require the system initially prepared in a canonical distribution and the validity of the detailed balance (DB) condition. As a more universal IFT, the entropy production identity (EPI) \cite{Seifert_2005} holds for arbitrary initial distributions even without the DB condition. It is worth mentioning that a DFT of heat has been discovered quite recently, where the distribution of the initial state in the phase space is required to be uniform rather than canonical \cite{Park_2014}.

Here come several fundamental questions: Why are there strict requirements for the distributions of the initial state in these FTs, and can they be released? Can these existing FTs be traced back to the same root? 
Besides the FTs of the work, the heat, and the trajectory entropy production, are there any other FTs associated with other thermodynamic variables? 
These questions were partially answered by the unified FTs explored by Seifert \cite{Seifert_2008,Seifert_2012}. For a stochastic system undergoing a nonequilibrium process within the time interval $[0,\tau]$, driven by a temporally varying work parameter $\lambda_t$ ($0 \leq t \leq \tau$), 
the unified IFT reads \cite{Seifert_2008,Seifert_2012}
\begin{equation}
\langle e^{-R} \rangle = \langle \frac{p^a_\tau (\Gamma_\tau)}{p_0 (\Gamma_0)}e^{- \Delta s_m} \rangle = 1 \; ,
\label{uniIFT}
\end{equation}
where the trajectory-dependent functional $R[\Gamma_t] \equiv \ln [ \mathcal{P} (\Gamma_t)/ \mathcal{\bar P} (\bar \Gamma_t)] = \Delta s_m - \ln \bar p_0 (\Gamma^\dagger_\tau) + \ln p_0 (\Gamma_0)$. Here 
$\Delta s_m$ is the entropy production in the medium \cite{Seifert_2005}; $p^a_\tau (\Gamma)$ is the auxiliary final distribution in the phase space; $p_0 (\Gamma)$ ($\bar p_0 (\Gamma)$) is the initial distribution of the system in the phase space for the forward (backward) process; $\mathcal{P} (\Gamma_t)$ ($\mathcal{\bar P} (\bar \Gamma_t)$) is the probability density of a trajectory $\Gamma_t$ ($\bar \Gamma_t$) in the trajectory space due to the protocol $\lambda_t$ (time-reverse protocol $\bar \lambda_t \equiv \lambda_{\tau - t}$); $\bar \Gamma_t \equiv \Gamma^\dagger_{\tau - t}$ is the time-reverse trajectory of $\Gamma_t$ with $\Gamma^\dagger$ to be the time-reversal of the phase space point $\Gamma$ (e.g., for the underdamped Langevin dynamics, $\Gamma_t = (\boldsymbol{r}_t,\boldsymbol{p}_t)$, then $\Gamma^\dagger_t = (\boldsymbol{r}_t,-\boldsymbol{p}_t)$ and $\bar \Gamma_t = (\boldsymbol{r}_{\tau - t},-\boldsymbol{p}_{\tau - t}$)). In fact, Eq.~(\ref{uniIFT}) is a unification of the IFT of heat, the JE and the EPI, since they can be respectively generated by setting the initial distribution $p_0 (\Gamma_0)$ and the auxiliary final distributions $p^a_\tau (\Gamma_\tau)$ to be both uniform, both canonical and the distributions connected by the real dynamic evolution. For those processes the thermodynamic variable $S_\alpha$ with a definite time-reversal parity, say $S_\alpha[\bar \Gamma_t] = \epsilon_\alpha S_\alpha[\Gamma_t]$, $\epsilon_\alpha = \pm 1$, we further have the following unified DFT \cite{Seifert_2012}
\begin{equation}
\frac{\bar P (\{S_\alpha = \epsilon_\alpha s_\alpha\})}{P (\{S_\alpha = s_\alpha\})} = \langle e^{-R} | \{S_\alpha = s_\alpha\} \rangle \; ,
\label{uniDFT}
\end{equation}
where $P (\{S_\alpha\})$($\bar P (\{S_\alpha\})$) denotes the joint distribution of multiple variables $S_\alpha$ for the forward (backward) process \cite{Kolton_2010,Kolton_2012}, and the right hand side (r. h. s.) means the conditional expectation of $e^{-R}$ when $S_\alpha = s_\alpha$. To generate the DFT of heat or the CFT, we simply choose the single odd parity quantity $S$ to be the heat ($Q$) or the work ($W$), then set $p_0 (\Gamma)$ and $\bar p_0 (\Gamma)$ to be both uniform distributions or both canonical distributions. This choice leads to $R = \beta Q$ or $\beta (W - \Delta F)$ and thus makes the r. h. s. of Eq.~(\ref{uniDFT}) simply $e^{-\beta Q}$ or $e^{-\beta (W-\Delta F)}$, since $R$ is uniquely determined by $S$ and the conditional distribution is a delta one. From this point of view, we can say that the FTs mentioned in the last paragraph do share the same root, which might be summarized as a combination of the microscopic reversibility (MR) and the dynamic property. The MR ensures the validity of Eq.~(\ref{uniIFT}) (and Eq.~(\ref{uniDFT}), though not obvious), since it is no more than the probability-normalization relation of all the time-reverse trajectories whose forward ones are of nonzero probability. In fact, the conventional FTs are inapplicable to absolutely irreversible processes because of the breaking down of the MR, as has been highlighted in recent investigations \cite{Ueda_2014}. The dynamic property, such as the DB condition, is necessary to relate certain thermodynamic variable to $R$ by properly choosing the two initial distributions. This is important for endowing specific physical meaning to the abstract identities (\ref{uniIFT}) and (\ref{uniDFT}) as merely the corollaries of the MR.

On the other hand, despite of their universal validity, the unified IFT (\ref{uniIFT}) will be physically meaningless if $R$ can't be related to certain thermodynamic variables, while the r. h. s. of the unified DFT (\ref{uniDFT}) is usually difficult to either calculate or be given a transparent physical interpretation, unless $R$ uniquely depends on $S$. Such entanglement between the thermodynamic variable and the initial distributions is the reason why the EPI holds for arbitrary distributions while the FTs of the heat and the work require specific initial distributions: the functional of entropy production naturally contains the distribution functions of the initial and final phase space points, but the heat and the work have nothing to do with them.

If the initial distribution is not canonical (uniform) distribution, can we still construct a fluctuation theorem for the work (heat)? According to the above analysis, 
the answer seems to be no. However, in this article, we propose a new refined unified FT which achieves the separation of the thermodynamic variable 
from the choices of the initial distributions. As a result of the DB condition, this refined unified FT is even more ``detailed" than DFT (\ref{uniDFT}), because it involves the joint distributions with the phase space point. In the studies of FTs, the first encounter of the joint distributions with the phase space point is in the Hummer-Szabo relation \cite{Jarzynski_1997b,Crooks_2000,Hummer_2001}. However, the Hummer-Szabo relation still requires the distribution of the initial state to be a canonical distribution, and is valid for the work only. In our current investigation, we extend the FTs of work and heats to an arbitrary initial distribution and even to other variables, such as the entropy production. The cost is that we need to know the joint distribution function with the phase space point, such as $P_{\tau}(W,\Gamma)$, which is more detailed than the usual distribution function, such as the work distribution $P_{\tau}(W)$. 
We show that the new refined unified FT can generate many existing FTs as well as many new FTs that are not known previously by choosing proper initial states. 
It may also be of potential values in investigating information thermodynamics, where the initial distributions might be quite irregular due to the extra distribution rectification by the information.

We notice that similar problems, i.e., the FTs for arbitrary initial states, are investigated in a recent work \cite{Esposito_2014}. However, the systems considered therein are discrete, automatous (not externally driven) and may not satisfy the DB condition, and the FTs are associated with the local currents in a general dynamic network. Thus both aspects of their study are complementary to ours. This issue is also discussed in Ref.~\cite{Lahiri_2014}, but is not its main focus.

This paper is organized as follows. In Sec.~\hyperref[sec:derivation]{II}, we derive the refined unified FT by analyzing its close relation to the DB condition for general stochastic thermodynamic systems. In Sec.~\hyperref[sec:relation]{III}, we reproduce the existing FTs and generate some new ones of the work, the heat and the entropy production as some examples of the refined unified FT. Some applications of the refined unified FT are explored in Sec.~\hyperref[sec:application]{IV}.  In Sec.~\hyperref[sec:conclusion]{V}, we summarize the paper. Six appendices are added at the end of the paper, which we think are helpful for understanding the details of the paper and the relevant issues.

\section{Main Result and its Derivation}
\label{sec:derivation}

\subsection{Main result --- the refined unified FT}
We focus on general stochastic thermodynamic systems with the DB condition. The system is coupled to a heat bath with the inverse temperature $\beta$, thus undergoes isothermal processes within the time interval $[0,\tau]$ due to an external driving protocol denoted by $\lambda_t$, $t \in [0,\tau]$. We consider a trajectory-dependent thermodynamic quantity $ A [\Gamma_t]$ in the following form
\begin{equation}
A[\Gamma_t] = \beta Q[\Gamma_t] + a_\tau (\Gamma_\tau) - a_0 (\Gamma_0) \; .
\label{Adef}
\end{equation}
Here the heat functional $Q[\Gamma_t] \equiv -\int_{0}^{\tau} dt \dot{\Gamma}_t \partial_{\Gamma} U_t (\Gamma_t )$ ($Q>0$ corresponds to the release of the heat from the system to the heat bath), with $U_t (\Gamma)$ to be the energy of phase space point $\Gamma$ at time $t$; $a_t (\Gamma)$ can be an arbitrary time-dependent function of the phase space point $\Gamma$, for which we can generally define its time-reversal $\bar a_t (\Gamma) \equiv a_{\tau - t} (\Gamma^\dagger )$ \cite{TRdef} and correspondingly
\begin{equation}
\bar A[\Gamma_t] = \beta Q[\Gamma_t] + \bar a_\tau (\Gamma_\tau) - \bar a_0 (\Gamma_0) \; .
\label{barAdef}
\end{equation}
All over the paper, if not particularly indicated, we always stipulate that the energy possesses the property of the time-reversal invariance, i.e., $U_t (\Gamma^\dagger) = U_t (\Gamma)$. Our main result reads
\begin{equation}
\begin{split}
\frac{\int_{\mathfrak{S}_0} d\Gamma \bar P_\tau (-A,\Gamma) p_0 (\Gamma^\dagger) e^{\bar a_\tau (\Gamma)}} {\int_{\mathfrak{S}_\tau} d\Gamma P_\tau (A,\Gamma) \bar p_0 (\Gamma^\dagger) e^{a_\tau (\Gamma)} } = e^{-A} \; ,
\end{split}
\label{mainres}
\end{equation}
where the accessible phase space at time $t$ ($\tau-t$) for the forward (backward) process is denoted by $\mathfrak{S}_t$, thus the integral of $\Gamma$ in the denominator (numerator) goes over the whole accessible phase space of $\Gamma_\tau$ ($\Gamma_0$ or $\bar \Gamma_\tau \equiv \Gamma^\dagger_0$) \cite{acphsp}; $p_0 (\Gamma)$ ($\bar p_0 (\Gamma)$) denotes the distribution of the initial state in the phase space for the forward (backward) process; $P_\tau (A,\Gamma)$ ($\bar P_\tau (\bar A,\Gamma)$) is the joint distribution function of the thermodynamic variable $ A $ ($\bar A$) accumulated until time $\tau$ and the final (at time $\tau$) phase space point, starting from the initial distribution $p_0 (\Gamma)$ ($\bar p_0 (\Gamma)$) and driven under the protocol $\lambda_t$ ($\bar \lambda_t \equiv \lambda_{\tau -t}$). Eq.~(\ref{mainres}) is valid for arbitrary initial distributions $p_0 (\Gamma)$ and $\bar p_0 (\Gamma)$, and arbitrary state functions $a_t (\Gamma)$. Also, similar to the JE and the CFT, it is valid for an arbitrary protocol $\lambda_t$ and an arbitrary driving time $\tau$.

\subsection{Dynamic property equivalent to the DB condition}
In order to demonstrate the close relation between our main result (\ref{mainres}) and the DB condition transparently, we use a somehow complicated method to carry out the derivation, though a relatively simple but much more mathematical proof is available by using the path integral approach (see Appendix~\hyperref[app:PIderivation]{A}). 

We first write down the Fokker-Planck equation (FPE) of the stochastic system
\begin{equation}
\partial_{t} p_t (\Gamma) = \mathcal{L}_t p_t (\Gamma) \; ,
\label{FPE}
\end{equation}
with the generator $\mathcal{L}_t$ to be a time-dependent linear operator corresponding to the protocol $\lambda_t$ and only acting on $\Gamma$. For later use, we define the transpose operator of $\mathcal{L}_t$ (denoted by $\mathcal{L}^T_t$), which satisfies
$\int_{\mathfrak{S}_t} d\Gamma g(\Gamma) \mathcal{L}_t f(\Gamma) = \int_{\mathfrak{S}_t} d\Gamma f(\Gamma) \mathcal{L}^T_t g(\Gamma)$
for arbitrary normalizable (i.e., $|\int d\Gamma f(\Gamma)| < +\infty$) functions $f(\Gamma)$ and $g(\Gamma)$ defined in the phase space. One can see that the normalization of $p_t (\Gamma)$ will impose the property $\mathcal{L}^T_t 1 = 0$ to $\mathcal{L}_t$ if we integrate the FPE (\ref{FPE}) over $\Gamma$, though this is not rigorous for the systems with infinite phase space where $1$ is not normalizable. 
We can also define the time-reversed operator of $\mathcal{L}_t$ (denoted by $\mathcal{L}^\dagger_t$). It is obtained by adding minus signs to all the components in $\mathcal{L}_t$ with the odd time-reversal parity (e.g., $\boldsymbol{p} \to -\boldsymbol{p}$, $\partial_{\boldsymbol{p}} \to - \partial_{\boldsymbol{p}}$,  $\boldsymbol{p}$ is momentum). 

In terms of the generator, the DB condition manifests itself in the following algebraic symmetry
\begin{equation}
e^{\beta U_t (\Gamma)} \mathcal{L}^\dagger_t e^{-\beta U_t (\Gamma)} = \mathcal{L}^T_t\;.
\label{crucialdp}
\end{equation}
Such dynamic property comes directly from the combination of the original definition of the DB condition $e^{-\beta U_t(\Gamma_1)}w_t(\Gamma_1\to\Gamma_2)=e^{-\beta U_t(\Gamma_2)}w_t(\Gamma^\dagger_2\to\Gamma^\dagger_1)$ (for any $\Gamma_1,\Gamma_2\in\mathfrak{S}_t$) and the transition rate formula $w_t(\Gamma_1\to\Gamma_2)=\int_{\mathfrak{S}_t } d\Gamma \delta (\Gamma - \Gamma_2) \mathcal{L}_t \delta (\Gamma - \Gamma_1)$ \cite{Risken_1989}. Conversely, the original DB condition follows if Eq.~(\ref{crucialdp}) is previously assumed, thus the dynamic property is actually equivalent to the DB condition. We emphasize that Eq.~(\ref{crucialdp}) should be valid at any time $t\in[0,\tau]$ despite the work parameter is temporally varied. This is the straightforward dynamic generalization of the common static version (where the work parameter is fixed), which can be found in the standard literature \cite{Risken_1989}. We also want to mention that Eq.~(\ref{crucialdp}) ensures when we suddenly stop the external driving at time $t$, the system will always relax to the canonical distribution determined by $\lambda_t$ owing to the property $\mathcal{L}^T_t1=0$. In addition, Eq.~(\ref{crucialdp}) is much more stronger than merely imposing $\mathcal{L}_t e^{-\beta U_t (\Gamma)}=0$, that's why the balance is called detailed.

As an example, it can be checked that Eq.~(\ref{crucialdp}) is true for the Langevin dynamics (in both the underdamped and the overdamped regimes, see Appendix~\hyperref[app:cdpforLD]{B}). It is also notable that a discrete but more general (may be without the DB condition) version of Eq.~(\ref{crucialdp}) has appeared in Ref.~\cite{Harris_2007} (see Eq. (3.24) therein).

\subsection{Derivation of Eq.~(\ref{mainres}) based on the characteristic function}
According to the FPE (\ref{FPE}) as well as the definition of the functional $A$ (\ref{Adef}), it can be proved (see Appendix~\hyperref[app:generalEOM]{C} for a more general formula) that the joint distribution function $P_t (A,\Gamma)$ satisfies the following equation of motion
\begin{equation}
\begin{split}
\partial_{t} P_t (A,\Gamma) = \{ &e^{-[a_t(\Gamma)-\beta U_t(\Gamma)] \partial_{A}} \mathcal{L}_t e^{[a_t(\Gamma)-\beta U_t(\Gamma)] \partial_{A}}\\ &- \partial_{t} a_t(\Gamma) \partial_{A} \} P_t (A,\Gamma) \; .
\end{split}
\label{EOM}
\end{equation}
Notice that the operator acting on $P_t (A,\Gamma)$ on the r. h. s. only contains $\partial_{A}$ but is independent of $ A $. It is natural to (partly) perform an integral transformation, as is always done when we construct the Feynman-Kac formula (e.g., $a_t (\Gamma)=\beta U_t (\Gamma)$ in Eq.~(\ref{EOM})). We take the inverse Fourier transformation to change $P_t (A,\Gamma)$ into its characteristic function $G_t (\mu,\Gamma) \equiv \int^{+ \infty}_{- \infty} dA e^{i \mu A} P_t (A,\Gamma)$. For further simplicity, we define the modified characteristic function $G^m_t (\mu,\Gamma) \equiv e^{-i \mu a_t(\Gamma)} G_t (\mu,\Gamma)$ . Based on Eq.~(\ref{EOM}), it is found that $G^m_t (\mu,\Gamma)$ satisfies the following equation of motion
\begin{equation}
\partial_{t} G^m_t (\mu,\Gamma) = \mathcal{L}_t (\mu) G^m_t (\mu,\Gamma) \; ,
\label{FEOM}
\end{equation}
where $\mathcal{L}_t (\mu) \equiv e^{-i \mu \beta U_t (\Gamma)} \mathcal{L}_t e^{i \mu \beta U_t (\Gamma)}$. For the backward process, we define $\mathcal{\bar L}_t (\mu) \equiv \mathcal{L}_{\tau - t} (\mu)$, so the equation of motion of the modified characteristic function of the backward process, $\bar G^m_t(\mu,\Gamma) \equiv e^{-i \mu \bar a_t(\Gamma)} \bar G_t(\mu,\Gamma)$, can be expressed as
\begin{equation}
\partial_{t} \bar G^m_t (\mu,\Gamma) = \mathcal{\bar L}_t (\mu) \bar G^m_t (\mu,\Gamma) \; .
\label{BEOM}
\end{equation}
In terms of $\mathcal{L}_t(\mu)$, 
 the relation (\ref{crucialdp}) can be rewritten as
\begin{equation}
\mathcal{L}^\dagger_t (\mu + i) = \mathcal{\bar L}^T_{\tau - t} (-\mu)
\label{dprw}
\end{equation}
by using the identity $\mathcal{L}^T_t (\mu) = e^{i \mu \beta U_t (\Gamma)} \mathcal{L}^T_t e^{-i \mu \beta U_t (\Gamma)}$. In order to make full use of Eq.~(\ref{dprw}), we write in the following special forms the equation of motions for both the forward (\ref{FEOM}) and the backward (\ref{BEOM}) processes
\begin{equation}
\partial_{t} G^m_t (\mu + i,\Gamma^\dagger) = \mathcal{L}^\dagger_t (\mu + i) G^m_t (\mu + i,\Gamma^\dagger) \; ,
\label{sFEOM}
\end{equation}
\begin{equation}
-\partial_{t} \bar G^m_{\tau - t} (-\mu,\Gamma) = \mathcal{\bar L}_{\tau - t} (-\mu) \bar G^m_{\tau - t} (-\mu,\Gamma) \;.
\label{sBEOM}
\end{equation}
Combining these two equations of motion with Eq.~(\ref{dprw}), after a straight forward calculation, we deduce that
\begin{equation}
\partial_{t} \int_{\mathfrak{S}_t} d\Gamma G^m_t (\mu + i,\Gamma^\dagger) \bar G^m_{\tau - t} (-\mu,\Gamma) = 0 \; .
\end{equation}
This means the integral $\int_{\mathfrak{S}_t} d\Gamma G^m_t (\mu + i,\Gamma^\dagger) \bar G^m_{\tau - t} (\mu,\Gamma)$ is a conserved quantity during the dynamic evolution. Particularly, at the terminal time points ($t=0,\tau$), we have 
\begin{equation}
\begin{split}
&\int_{\mathfrak{S}_\tau} d\Gamma G^m_\tau (\mu + i,\Gamma) \bar G^m_0 (-\mu,\Gamma^\dagger) \\
= &\int_{\mathfrak{S}_0} d\Gamma G^m_0 (\mu + i,\Gamma^\dagger) \bar G^m_\tau (-\mu,\Gamma) \; .
\end{split}
\label{terminal}
\end{equation}
After substituting the expressions of the modified characteristic functions into Eq.~(\ref{terminal}), we obtain 
\begin{equation}
\begin{split}
&\int_{\mathfrak{S}_\tau} d\Gamma G_\tau (\mu + i,\Gamma) \bar p_0 (\Gamma^\dagger) e^{a_\tau (\Gamma)} \\
= &\int_{\mathfrak{S}_0} d\Gamma \bar G_\tau (-\mu,\Gamma) p_0 (\Gamma^\dagger) e^{\bar a_\tau (\Gamma)} \; .
\end{split}
\label{mainresCF}
\end{equation}
By taking the Fourier transformation of Eq.~(\ref{mainresCF}), we finally obtain our main result (\ref{mainres}).

It is worth mentioning that a similar method based on the symmetry of the generator has appeared in a recent work \cite{Virtanen_2014} (see Eq.~(\ref{dprw}) here and Eq.~(10) therein). But their focus was on the Bochkov-Kuzovlev equality \cite{Liu_2014a} for the open quantum systems described by Lindblad master equations. Therefore, we believe there is a quantum generalization of the main result (\ref{mainres}) (and its extensive corollaries) for at least  Lindblad-type open quantum systems, which we leave for our future work.

\section{Refinement of the Existing FTs and Generating of New FTs}
\label{sec:relation}

\subsection{refined FT of work}
The simplest specialization of the refined unified FTs (\ref{mainres}) is when we choose $a_t (\Gamma) = \beta U_t (\Gamma)$. In this manner, both the quantity $A$ (\ref{Adef}) and $\bar A$ (\ref{barAdef}) turn out to be the dimensionless work $\beta W$, due to the first law of thermodynamics at the level of individual trajectories \cite{Sekimoto_1998}. The refined FT of work reads
\begin{equation}
\begin{split}
\frac{\int_{\mathfrak{S}_0} d\Gamma \bar P_\tau (-W,\Gamma) p_0 (\Gamma^\dagger) e^{\beta U_0 ( \Gamma^\dagger)}} {\int_{\mathfrak{S}_\tau} d\Gamma P_\tau (W,\Gamma) \bar p_0 (\Gamma^\dagger) e^{\beta \bar U_0 (\Gamma^\dagger)}} = e^{-\beta W}  \; .
\end{split}
\label{workFT}
\end{equation}
Similar to the CFT, this relation is valid for an arbitrary driving protocol $\lambda_{t}$ 
and an arbitrary driving time $\tau$. What is more, this relation is more general than the CFT because it is valid for arbitrary initial distributions $p_0 (\Gamma)$ and $\bar p_0 (\Gamma)$. Obviously, if we want to construct the existing work FTs from Eq.~(\ref{workFT}), the choices of the distributions of the initial state in the phase space should be the canonical ones for both the forward and the backward processes. That is, $p_0(\Gamma)=p^{eq}_0 (\Gamma)\equiv e^{-\beta U_0 (\Gamma)}/Z_0 (\beta)$ and $\bar p_0(\Gamma)=\bar p^{eq}_0 (\Gamma)\equiv e^{-\beta \bar U_0 (\Gamma)}/ \bar Z_0 (\beta)$, where the partition functions $Z_0 (\beta) \equiv \int_{\mathfrak{S}_0} d\Gamma e^{-\beta U_0 (\Gamma)}$ and $\bar Z_0 (\beta) \equiv \int_{\mathfrak{S}_\tau} d\Gamma e^{-\beta \bar U_0 (\Gamma)}$. For such choices, Eq.~(\ref{workFT}) reduces to the well-known CFT \cite{Crooks_1999}
\begin{equation}
  \frac{\bar P_\tau (-W)}{P_\tau (W)} = e^{-\beta (W - \Delta F)}\; ,
\label{CFT}
\end{equation}
where $P_\tau (W) \equiv \int_{\mathfrak{S}_\tau} d\Gamma P_\tau (W,\Gamma)$ and $\bar P_\tau (W) \equiv \int_{\mathfrak{S}_0} d\Gamma \bar P_\tau (W,\Gamma)$, $\Delta F \equiv - \beta^{-1} \ln [Z_\tau (\beta)/ Z_0 (\beta)]$ is the free energy difference \cite{deltaF}. The integral version of the CFT (\ref{CFT}) is the celebrated JE \cite{Jarzynski_1997}
\begin{equation}
\langle e^{-\beta (W - \Delta F)} \rangle = 1 \; .
\label{JE}
\end{equation}
We can also easily reproduce the Hummer-Szabo relation \cite{Hummer_2001}
\begin{equation}
\langle \delta (\Gamma_\tau - \Gamma') e^{-\beta W} \rangle = \frac{e^{-\beta U_\tau (\Gamma')}}{Z_0(\beta)} \; ,
\label{HSR}
\end{equation}
as long as we set the initial distribution for the forward process to be a canonical distribution $p^{eq}_0 (\Gamma)$ and the initial distribution for the backward process to be a $\delta$ distribution $\bar p_0 (\Gamma) = \delta (\Gamma-\Gamma'^\dagger)$ in Eq.~(\ref{workFT}).

The discrete version of Eq.~(\ref{workFT}) is
\begin{equation}
\frac{\sum_{\bar m} \bar P_\tau (-W,\bar m) p_0 (\bar m^\dagger) e^{\beta \bar E^{\bar m}_\tau}}{\sum_n P_\tau (W,n) \bar p_0 (n^\dagger) e^{\beta E^n_\tau} }= e^{-\beta W}  \; ,
\label{dworkFT}
\end{equation}
where $E^n_t$ ($\bar E^{\bar m}_t$) is the energy of the state $n$ ($\bar m$) at time $t$ for the forward (backward) process; $n^\dagger$ denotes the time-reversed state of $n$, e.g., the spin-down state if $n$ denotes spin-up. Certainly, this relation holds for discrete-level stochastic systems. We would like to emphasize that Eq.~(\ref{dworkFT}) is also valid for an isolated quantum system (not necessarily with time-reversal symmetry), where the initial density matrix is generally $\varrho_0 = \sum_m p_0 (m)| m \rangle \langle m |$ (and $\bar \varrho_0 = \sum_{\bar n} \bar p_0 (\bar n) | \bar n \rangle \langle \bar n |$ for the time-reversed process), as long as we admit the two-point projection measurement definition of quantum work (see Appendix~\hyperref[app:qworkFT]{D}). Furthermore, $\beta$ in Eq.~(\ref{dworkFT}) can be arbitrarily chosen and may even be a complex number, since the quantum system is isolated from any heat bath.

\subsection{generating new FTs of work}
In addition, from the refined FT of work (\ref{workFT}) one can derive several FTs of work that were previously not known to researchers in this field. For that purpose, let us first choose both $p_0 (\Gamma)=\delta(\Gamma-\Gamma_0)$ and $\bar p_0 (\Gamma)=\delta(\Gamma-\Gamma'^\dagger)$ to be $\delta$ distributions. So that Eq.~(\ref{workFT}) becomes 
\begin{equation}
\begin{split}
P_\tau (W,\Gamma'|\Gamma_0)e^{-\beta (W-\Delta F)}=\frac{\bar p^{eq}_{0}(\Gamma'^{\dagger})}{p^{eq}_{0}(\Gamma_0)} \bar P_\tau (-W,\Gamma_0^{\dagger}|\Gamma'^{\dagger})\;,
\end{split}
\label{conditionaljointdistribution}
\end{equation}
where $P_\tau (W,\Gamma'|\Gamma_0)$ is the conditional joint probability distribution. It means the sum of the probabilities of all the forward trajectories that end at $\Gamma'$, and the work accumulated along each of these trajectories are equal to $W$ conditioned on the given initial phase space point $\Gamma_0$. $\bar P_\tau (-W,\Gamma_0^{\dagger}|\Gamma'^{\dagger})$ can be understood in a similar way but for the backward process. This relation can be regarded as a generalization of CFT to initial $\delta$ distributions, and has previously been obtained in Refs.~\cite{Jarzynski_2000,Karplus_2008}. We do integral with respect to $W$ on both sides of Eq.~(\ref{conditionaljointdistribution}) and obtain
\begin{equation}
\begin{split}
\left\langle \delta(\Gamma_{\tau}-\Gamma') \left. e^{-\beta (W-\Delta F)} \right|\Gamma_0 \right\rangle= \frac{\bar p^{eq}_{0}(\Gamma'^{\dagger})}{p^{eq}_{0}(\Gamma_0)} \bar p_\tau (\Gamma^{\dagger}_0|\Gamma'^{\dagger})\;,
\end{split}
\label{generalizedhummerszabo}
\end{equation}
where $\left\langle F[\Gamma_t]|\Gamma_0 \right\rangle$ indicates the ensemble average of the functional $F[\Gamma_t]$ over all trajectories that start from the phase space point $\Gamma_0$, and $\bar p_\tau (\Gamma^{\dagger}_0|\Gamma'^{\dagger})$ is the conditional probability distribution in the phase space for the backward process. It describes the final probability distribution of the backward process at $\Gamma^{\dagger}_0$ given the initial state at $\Gamma'^{\dagger}$. This relation (\ref{generalizedhummerszabo}) can be regarded as a generalization of the Hummer-Szabo relation (\ref{HSR}) to the $\delta$ initial distribution. If we further do integral with respect to $\Gamma'$ on both sides of Eq.~(\ref{generalizedhummerszabo}), we obtain a JE-like FT
\begin{equation}
\left\langle \left. e^{-\beta (W-\Delta F)} \right|\Gamma_0 \right\rangle= \frac{\bar p_\tau (\Gamma^\dagger_0)}{p^{eq}_{0}(\Gamma_0)} \;,
\label{JEforarbitrary}
\end{equation}
where $\bar p_\tau (\Gamma)$ is the final probability distribution in the phase space evolved from the initial canonical distribution $\bar p^{eq}_{0}(\Gamma)$ in the time-reversed process. This relation can be regarded as the generalization of JE to the $\delta$ initial distribution. Similar mathematical relations have been obtained in Ref.~\cite{Kawai_2007} for isolated classical systems and in Ref.~\cite{Liu_2014} for open quantum systems. But in both these two references, their focus are on different problems, and the mathematical relations are not interpreted in this way.

Besides the initial $\delta$ distribution, the JE (\ref{JE}) and the Hummer-Szabo relation (\ref{HSR}) can actually be extended to an arbitrary initial distribution in the forward process. We multiply an arbitrary initial distribution $p_{0}(\Gamma_0)$ to both sides of Eq.~(\ref{conditionaljointdistribution}) before we do the integral with respect to $\Gamma_0$, and we obtain the extended Hummer-Szabo relation for an arbitrary initial distribution $p_{0}(\Gamma)$ 
\begin{equation}
\begin{split}
&\left\langle \delta(\Gamma_{\tau}-\Gamma') e^{-\beta (W-\Delta F)} \right\rangle_{p_{0}(\Gamma)}\\
=&\int_{\mathfrak{S}_0} d \Gamma \bar p_\tau ( \Gamma^\dagger|\Gamma'^{\dagger})\bar p^{eq}_{0}(\Gamma'^{\dagger}) \frac{p_{0}(\Gamma) }{p^{eq}_{0}(\Gamma)} \;,
\end{split}
\label{HZforarbitrary}
\end{equation}
where $\left\langle ...\right\rangle_{p_{0}(\Gamma)}$ represents the average over all trajectories when the initial distribution is $p_{0}(\Gamma)$ for the forward process. If we further do integral with respect to $\Gamma'$ on both sides of Eq.~(\ref{HZforarbitrary}) we obtain the generalized JE for an arbitrary initial distribution 
\begin{equation}
\begin{split}
\left\langle e^{-\beta (W-\Delta F)} \right\rangle_{p_{0}(\Gamma)}=\int_{\mathfrak{S}_0} d \Gamma \bar p_\tau (\Gamma^\dagger) \frac{p_{0}(\Gamma)}{p^{eq}_{0}(\Gamma)}\;.
\end{split}
\label{Wexpavg}
\end{equation}
In this equality, the initial distribution of the forward process can be an arbitrary distribution \textcolor{blue}{$p_{0}(\Gamma)$}, while the initial distribution of the backward process must be a canonical distribution. It is worth mentioning that in the FT of the total entropy production along individual trajectories \cite{Seifert_2005}, the initial distribution can be an arbitrary distribution (\ref{EPI}). It was assumed previously that only when the initial distribution is a canonical distribution (globally thermal equilibrium) or a partially thermal equilibrium distribution \cite{Ritort_2009,Ritort_2012} can one construct the FT of work. Here we generalize the FT of work (mainly JE (\ref{JE}), CFT (\ref{CFT}) and Hummer-Szabo relation (\ref{HSR})) to an arbitrary initial state. The generalized JE for an arbitrary initial state (\ref{Wexpavg}) may have potential applications in free energy recovering via nonequilibrium work measurement and numerical free energy computation.

\subsection{refined FT of heat}
Another specialization of the refined unified FT (\ref{mainres}) is the FT of heat when $a_t (\Gamma)$ equals to a constant (e.g., zero), thus $A =\bar A=\beta Q$. The refined FT of heat reads 
\begin{equation}
\frac{\int_{\mathfrak{S}_0} d\Gamma \bar P_\tau (-Q,\Gamma) p_0 (\Gamma^\dagger)}{\int_{\mathfrak{S}_\tau} d\Gamma P_\tau (Q,\Gamma) \bar p_0 (\Gamma^\dagger)}  = e^{-\beta Q} \; .
\label{heatFT}
\end{equation}
Similar to the DFT of heat \cite{Park_2014}, this relation is valid for an arbitrary driving protocol
$\lambda_{t}$ and driving time $\tau$.
Moreover, this relation is more general than the DFT of heat, because it is valid for arbitrary initial distributions $p_0 (\Gamma)$ and $\bar p_0 (\Gamma)$. If we want to reduce the joint distribution function to the heat distribution function, the only choice is to make both $p_0 (\Gamma)$ and $\bar p_0 (\Gamma)$ independent of the phase space point, i.e., the uniform distribution, 
as was found in Ref.~\cite{Park_2014}. However, a uniform distribution can never be truly realized in continuous systems, where the entropy has no upper bound due to the infinite volume of the phase space. So we have to consider finite-level systems, such as spin systems. Similar to Eq.~(\ref{dworkFT}) , the discrete version of Eq.~(\ref{heatFT}) is
\begin{equation}
\frac{\sum_{\bar m} \bar P_\tau (-Q,\bar m) p_0 (\bar m^\dagger)}{\sum_n P_\tau (Q,n) \bar p_0 (n^\dagger)}  = e^{-\beta Q} \; .
\label{dheatFT}
\end{equation}
Suppose that the system has totally $N$ states. By setting $\bar p_0 (n^\dagger) = p_0 (\bar m^\dagger) \equiv 1/N$ (maximum entropy state) in Eq.~(\ref{dheatFT}), we obtain 
\begin{equation}
\frac{\bar P_\tau (-Q)}{P_\tau (Q)}  = e^{-\beta Q} \; ,
\label{heatDFT}
\end{equation}
where $P_\tau (Q) \equiv \sum_n P_\tau (Q,n)$ and $\bar P_\tau (Q) \equiv \sum_{\bar m} \bar P_\tau (Q,\bar m)$ are respectively the heat distribution functions for the forward and the backward processes. Eq.~(\ref{heatDFT}) is nothing but the DFT of heat, whose integral version reads \cite{Park_2014}
\begin{equation}
\langle e^{-\beta Q} \rangle = 1 \; .
\label{heatIFT}
\end{equation}
It is worth mentioning that Eqs.~(\ref{heatDFT}) and (\ref{heatIFT}) are likely to be experimentally tested in analogy to the verification of the EPI for a finite-level system \cite{Schuler_2005}. More generally, we can test Eq.~(\ref{heatFT}) in whatever systems with finite phase space.

One may imagine that the uniform initial distribution of the DFT or the IFT of heat might be approached via the limit $\beta' \to 0$ ($T' \to \infty$), with $\beta'$ to be the inverse temperature of the initial canonical distribution. Hence, one may ask whether
\begin{equation}
\lim_{\beta' \to 0} \frac{1}{\beta Q} \ln \frac{P_\tau (Q)}{\bar P_\tau (-Q)} = 1
\label{lheatDFT}
\end{equation}
holds true even for a continuous system with its initial distribution to be $p_0 (\Gamma) = e^{-\beta' U_0 (\Gamma)}/Z_0(\beta')$. In fact, it has been demonstrated~\cite{Park_2014} that for a driven Brownian harmonic oscillator, if we rewrite $Q$ in $w \tau p$ , with $w$ to be a positive quantity that characterizes the driving speed, and take the limit $\tau \to \infty$ (so that $Q \to \infty$ for finite $p$) before $\beta' \to 0$, Eq.~(\ref{lheatDFT}) will be indeed true. It might be an intriguing but involved subject to investigate the validity range of Eq.~(\ref{lheatDFT}) for more general kinds of stochastic systems with infinite phase space.

\subsection{generating new FTs of heat}
Similar to JE for an arbitrary initial distribution, we can extend the integral FT of heat (\ref{heatIFT}) to an arbitrary initial state. For the sake of a well-defined uniform distribution, we write down the discrete version for an $N$-state system
\begin{equation}
\langle e^{-\beta Q} \rangle_{p_{0}(m)} = \sum_m \bar p_\tau (m^\dagger) \frac{p_0 (m)}{p^u_0 (m)} \; ,
\label{Qexpavg}
\end{equation}
where $p^u_0 (m)\equiv1/N$ is the initial uniform distribution of the forward process, and the final distribution of the microscopic state for the backward process $\bar p_\tau (m^\dagger)= \int dQ \bar P_\tau (Q,m^\dagger)$ in Eq.~(\ref{Qexpavg}) must correspond to the uniform 
initial distribution. One can see that the r. h. s. of Eq.~(\ref{Qexpavg}) usually deviates from $1$, if the initial distribution of the forward process is not prepared in the uniform distribution. Eq.~(\ref{Qexpavg}) is valid for an arbitrary protocol and an arbitrary 
driving time. What is more it does not require the initial distribution of the forward process to be the uniform distribution. Hence, if Eq.~(\ref{heatFT}) can be regarded as the generalizations of the detailed version of the heat FT \cite{Park_2014} to an arbitrary initial distribution, Eq.~(\ref{Qexpavg}) can be regarded as the generalization of the integral version of heat FT \cite{Park_2014} to an arbitrary initial distribution.

For the heat FT, one can also derive a Hummer-Szabo like relation. The initial distribution for the forward process and the reverse process are chosen to be the uniform distribution and the $\delta$ distribution respectively. Let $m_\tau$ be the final (at time $\tau$) state of the forward process, then this relation reads
\begin{equation}
\langle \delta_{m_\tau n} e^{-\beta Q} \rangle = \frac{1}{N} \; .
\label{heatHSR}
\end{equation}
Furthermore, we can also extend this Hummer-Szabo like relation for heat (\ref{heatHSR}) to an arbitrary initial distribution $p_0(m)$
\begin{equation}
\begin{split}
&\left\langle \delta_{m_\tau n} e^{-\beta Q} \right\rangle_{p_{0}(m)}
=\sum_m p_{0}(m) \bar p_\tau (m^\dagger|n^{\dagger})\;.
\end{split}
\label{heatHZforarbitrarysimple}
\end{equation}

Up to now, we have successfully generalized the FTs of work and heat to arbitrary initial distributions. For a comparison between previous results and our current results, please see Table \ref{table1}.

\subsection{Unified IFT and EPI}
Besides reproducing the JE, the CFT and the heat FTs, the refined unified FT (\ref{mainres}) can reproduce the EPI. Let's again have a look at the main result (\ref{mainres}). 
We find that if $a_\tau (\Gamma) = - \ln \bar p_0 (\Gamma^\dagger)$ and $\bar a_\tau (\Gamma) = - \ln p_0 (\Gamma^\dagger)$, which are sufficient to determine $a_0 (\Gamma) = a_\tau (\Gamma^\dagger) = - \ln p_0 (\Gamma)$ and $\bar a_0 (\Gamma) = a_\tau (\Gamma^\dagger) = - \ln \bar p_0 (\Gamma)$ (thus $ A [\Gamma_t]$ and $\bar A [\bar \Gamma_t]$ are completely determined) due to the definition of the time-reversal $\bar a_t (\Gamma) = a_{\tau - t} (\Gamma^\dagger)$, Eq.~(\ref{mainres}) will simply reduce to
\begin{equation}
\frac{ \bar P_\tau (-A)}{P_\tau (A)}= e^{-A}  \; .
\label{ADFT}
\end{equation}
Here $P_\tau (A) \equiv \int_{\mathfrak{S}_\tau} d\Gamma P_\tau (A,\Gamma)$ and $\bar P_\tau (A) \equiv \int_{\mathfrak{S}_0} d\Gamma \bar P_\tau (A,\Gamma)$. The quantity $ A [\Gamma_t] = \beta Q [\Gamma_t] - \ln \bar p_0 (\Gamma^\dagger_\tau) + \ln p_0 (\Gamma_0)$ depends on the initial distributions of both the forward and the backward processes, which is quite different from the cases of the heat and the work. The integral version of Eq.~(\ref{ADFT}) reads
\begin{equation}
\langle e^{-A} \rangle = \langle \frac{p^a_\tau (\Gamma_\tau)}{p_0 (\Gamma_0)}e^{-\beta Q} \rangle = 1 \;,
\label{AIFT}
\end{equation}
where the auxiliary final distribution in the phase space $p^a_\tau (\Gamma) \equiv \bar p_0 (\Gamma^\dagger)$ can be an arbitrary one, since the choice of $\bar p_0 (\Gamma)$ has no restriction. The generalized version of Eq.~(\ref{AIFT}) to the cases without the DB condition, which is merely to replace $\beta Q$ with $\Delta s_m$, is the unified IFT (\ref{uniIFT}) we mentioned in the beginning.

Despite of the fact that the quantity $A$ defined in such way always satisfies the DFT (\ref{ADFT}), usually it cannot be related to any thermodynamic variable we are familiar with. 
To see this, we consider the average of $A$ over all the trajectories
\begin{equation}
\langle A \rangle = \beta \langle Q \rangle + D[p_\tau(\Gamma) || \bar p_0(\Gamma^\dagger)] + \langle s \rangle_\tau - \langle s \rangle_0 \; ,
\label{Aavg}
\end{equation}
where $D[p_1(\Gamma) || p_2(\Gamma)] \equiv \int d\Gamma p_1 (\Gamma) \ln [p_1 (\Gamma) / p_2 (\Gamma)]$ is the Kullback-Leibler divergence between two probability distributions $p_1 (\Gamma)$ and $p_2 (\Gamma)$; $\langle s \rangle_t$ is the ensemble-average entropy of the system at time $t$; $p_\tau (\Gamma)$ is the final distribution in the phase space determined by the real dynamic evolution. A special case in which $A$ has a transparent physical meaning is when $\bar p_0 (\Gamma^\dagger) = p_\tau (\Gamma)$. In this case the second term in Eq.~(\ref{Aavg}) vanishes. For such a choice of the initial distribution for the backward process, $A$ is the trajectory-dependent total entropy production $\Delta s_{tot}$ for the forward process, and $\langle A \rangle$ is in consistency with the ensemble-average value $\langle \Delta s_{tot} \rangle$. However, since $\bar a_\tau (\Gamma)$ has been determined by $p_0 (\Gamma)$ as $- \ln p_0 (\Gamma^\dagger)$, $p_0 (\Gamma^\dagger)$ is usually different from the final phase space point distribution $\bar p_\tau (\Gamma)$ starting from the initial distribution $\bar p_0 (\Gamma) = p_\tau (\Gamma^\dagger)$ and driven by the time-reversed protocol \cite{Thermalization}. As a result, the functional $\bar A [\bar \Gamma_t] = \beta Q[\bar \Gamma_t] - \ln p_0 (\bar \Gamma^\dagger_\tau) + \ln \bar p_0 (\bar \Gamma_0)$ cannot be regarded as the total entropy production of the time-reversed trajectory $\bar \Gamma_t$. Hence, even when we choose $\bar p_0 (\Gamma^\dagger) = p_\tau (\Gamma)$, Eq.~(\ref{ADFT}) cannot be regarded as a detailed EPI. On the other hand, $\bar P_\tau (A)$ is still a normalized distribution function. So we can 
obtain the EPI \cite{Seifert_2005} by setting $A$ to be $\Delta s_{tot}$ without any problem
\begin{equation}
\langle e^{-\Delta s_{tot}} \rangle = 1 \; .
\label{EPI}
\end{equation}

Nevertheless, what if $p_0 (\Gamma^\dagger)$ happens to be the final distribution of the time-reversed process? In fact, such kind of specific distribution, named by echo state, has been recently studied by Van den Broeck and collaborators \cite{Van den Broeck_2015}, which clarifies the initial conditions of both the detailed EPI and the other DFTs in Ref.~\cite{Esposito_2010}. Similar to the discrete case that Van den Broeck et al discussed, for any given FPE (\ref{FPE}), the corresponding echo state $p^{echo}_0 (\Gamma)$ for the detailed EPI can be uniquely determined by solving
\begin{equation}
p^{echo}_0 (\Gamma) = \mathcal{T} \{ e^{\int^\tau_0 dt \mathcal{\bar L}^\dagger_t} \} \mathcal{T} \{ e^{\int^\tau_0 dt \mathcal{L}_t} \} p^{echo}_0 (\Gamma) \; ,
\end{equation}
where $\mathcal{T} \{ ... \}$ denotes the time-ordered product. For such choice of the initial distribution, the detailed EPI
\begin{equation}
 \frac{ \bar P_\tau (-\Delta s_{tot})}{P_\tau (\Delta s_{tot})} = e^{-\Delta s_{tot}}
\label{epDFT}
\end{equation}
holds unambiguously. This can be regarded as a generalization of the detailed EPI for steady state \cite{Seifert_2005}, where $\mathcal{L}_t$ should be time-independent, i.e., the work parameter should be fixed.

\section{Applications of the Refined Unified FT}
\label{sec:application}

\subsection{Calculating distribution functions of work and heat for arbitrary initial states}
Let's come back to the main result (\ref{mainres}). If we only set the initial distribution of the backward process to be $\bar p_0 (\Gamma) = e^{-\bar a_0 (\Gamma)} / \int_{\mathfrak{S}_\tau} d\Gamma e^{-\bar a_0 (\Gamma)}$, we will obtain
\begin{equation}
P_\tau (A) = e^A \int_{\mathfrak{S}_\tau} d\Gamma e^{-\bar a_0 (\Gamma)} \int_{\mathfrak{S}_0} d\Gamma \bar P_\tau (-A,\Gamma) p_0 (\Gamma^\dagger) e^{\bar a_\tau (\Gamma)} \; .
\label{PDFcal}
\end{equation}
This equation implies that $P_\tau (A)$ can be easily computed for arbitrary initial distribution $p_0 (\Gamma)$ as long as we know the joint distribution function $\bar P_\tau (A,\Gamma)$ corresponding to the particular initial distribution $\bar p_0 (\Gamma)$ for the time-reversed process. Specially, if $p_0 (\Gamma)$ is a delta function, Eq.~(\ref{PDFcal}) will become
\begin{equation}
P_\tau (A |\Gamma_0) = e^{A + a_0 (\Gamma_0)} \bar P_\tau (-A,\Gamma^\dagger_0) \int_{\mathfrak{S}_\tau} d\Gamma e^{-\bar a_0 (\Gamma)} \; ,
\label{deltaPDFcal}
\end{equation}
where $P_\tau (A | \Gamma_0)$ denotes the distribution of $A$ in condition that the initial state is known to be $\Gamma_0$. In fact, Eq.~(\ref{PDFcal}) and Eq.~(\ref{deltaPDFcal}) are equivalent to each other, similar to the equivalence between the dynamic property (\ref{crucialdp}) and the DB condition, due to the additivity of the probability for exclusive events, $P_\tau (A) = \int_{\mathfrak{S}_0} d\Gamma P_\tau (A | \Gamma) p_0 (\Gamma)$. As two examples, we can calculate the work and the heat distributions for any initial distributions $p_0 (\Gamma)$ ($p_0(m)$) via the following two formulas
\begin{equation}
P_\tau (W) = e^{\beta (W - \Delta F)} \int_{\mathfrak{S}_0} d\Gamma \bar P_\tau (-W,\Gamma^\dagger) \frac{p_0 (\Gamma)}{p^{eq}_0 (\Gamma)} \; ,
\label{WPDFcal}
\end{equation}
\begin{equation}
P_\tau (Q) = e^{\beta Q} \sum_m \bar P_\tau (-Q,m^\dagger) \frac{p_0 (m)}{p^u_0 (m)} \; .
\label{QPDFcal}
\end{equation}
These relations provide an alternative approach to obtain the work (heat) statistics besides, e.g., directly solving the Feynman-Kac formula for the forward process with an arbitrary initial condition. Such approach may be advantageous in certain cases due to the specific initial condition of the backward process.

As an example, let's consider an overdamped breathing Brownian harmonic oscillator in one dimension, of which the generator in the FPE reads
\begin{equation}
\mathcal{L}_t=\frac{1}{\gamma}\partial_x(k_t x+\beta^{-1}\partial_x) \; ,
\end{equation}
with the protocol chosen to be $k_t=k_0+\kappa t$. According to Ref.~\cite{Speck_2011}, using the Gaussian ansatz, the Feynman-Kac formula (partial differential equation) of this model can be reduced to a set of first-order ordinary differential equations, which can be easily solved numerically. Furthermore, if the initial state is the equilibrium state and the driving speed is slow, we can even perturbatively work out some analytical results, such as the mean work correction in the linear response regime. However, the direct perturbative analysis breaks down if the initial state deviates significantly from the equilibrium one. To bypass the difficulty, we make use the following equivalent form of Eq.~(\ref{WPDFcal})
\begin{equation}
G_\tau (\mu) = e^{-\beta\Delta F} \int_{\mathfrak{S}_0} d\Gamma \bar G_\tau (-\mu+i\beta,\Gamma^\dagger) \frac{p_0 (\Gamma)}{p^{eq}_0 (\Gamma)} \; ,
\label{WGFcal}
\end{equation}
where $\bar G_\tau(\mu,\Gamma)$ has an exact perturbative solution owing to its equilibrium initial state. In particular, if we choose $p_0(x)=(\beta' k_0/2\pi)^{\frac{1}{2}}e^{-\beta'k_0x^2/2}$, the canonical distribution of another temperature $\beta'^{-1}$, using the conclusions in Ref.~\cite{Speck_2011}, we will obtain a Gaussian distribution in the slow driving limit, of which the mean reads (see Appendix~\hyperref[app:Brownianoscillator]{E} for details)
\begin{equation}
\langle W\rangle = \Delta F +\frac{\gamma}{8\beta}[(\frac{2\beta}{\beta'}-1)\frac{1}{k^2_0}-\frac{1}{k^2_\tau}]\kappa+O(\kappa^2) \; ,
\label{meanwork}
\end{equation}
while the variance $\sigma^2_W$ is the same as that for the equilibrium initial state. Here the free energy difference $\Delta F=\ln\sqrt{k_\tau/k_0}$ and $k_\tau=k_0+\kappa\tau$. A notable inference follows that for an expanding process ($k_\tau>k_0$), $\langle W\rangle$ will be less than $\Delta F$ when $\beta'^{-1} < (k^2_0/k^2_\tau+1)\beta^{-1}/2$, which implies the necessity of an equilibrium initial state (with the same temperature of the heat bath) for the validity of the maximum work principle. Actually, the above analysis can be generalized to arbitrary slowly driven overdamped Langevin systems, whose Feynman-Kac formulas have a unified analytical treatment \cite{Speck_2004}.

\begin{table*}[tbp]
\caption{Comparison between previous results and our results on the requirements on the initial state for different FTs. Please note that in previous results only distribution functions of certain thermodynamic variables, such as $P_\tau(W)$ and $P_\tau(Q)$, are required. But in our current results, more detailed joint distribution functions, such as $P_\tau(W,\Gamma)$ and $P_\tau(Q,\Gamma)$, are required.}
\begin{center}
\begin{tabular}{|c|c|c|c|c|}
\hline
\multicolumn{2}{|c|} {Fluctuation Theorems}  & {\parbox{4.2cm}{Previous requirements on the distributions of initial state}} &  {\parbox{4cm}{Our requirements on the distributions of initial state}} \\
\hline
\multirow{3}{8em} {Work FTs}  & CFT \cite{Crooks_1999} & \parbox{5cm}{ canonical distribution (\ref{CFT})} & \parbox{4cm}{arbitrary distribution (\ref{workFT})} \\
& {JE \cite{Jarzynski_1997}} & \parbox{5cm}{ canonical distribution (\ref{JE})}  & \parbox{4cm}{ arbitrary distribution (\ref{Wexpavg})} \\
& Hummer-Szabo relation \cite{Hummer_2001} & \parbox{5cm}{ canonical distribution (\ref{HSR})} & \parbox{4cm}{arbitrary distribution (\ref{HZforarbitrary})}  \\
\hline
\multirow{3}{8em} {Heat FTs}& Detailed heat FT \cite{Park_2014} &  \parbox{5cm}{ uniform distribution (\ref{heatDFT})}  & \parbox{4cm}{ arbitrary distribution (\ref{heatFT})} \\
& Integral heat FT \cite{Park_2014} & \parbox{5cm}{ uniform distribution (\ref{heatIFT})}   & \parbox{4cm}{ arbitrary distribution (\ref{Qexpavg})}  \\
&Hummer-Szabo like relation  & N/A & arbitrary distribution (\ref{heatHZforarbitrarysimple}) \\
\hline
{FTs with feedback control} & Sagawa-Ueda equalities \cite{Sagawa_2010,Sagawa_2012} & \parbox{5cm}{ canonical distribution (\ref{SUE}),(\ref{SUE2})  arbitrary distribution (\ref{SUE3})}  & \parbox{4cm}{arbitrary distribution (\ref{gEPI})}  \\
\hline
\end{tabular}%
\end{center}
\label{table1}
\end{table*}

\subsection{Deriving FTs for feedback control processes}
Although the canonical distribution is much more common than other ones in real cases, the initial distribution corresponding to a specific protocol can be rather irregular in feedback control processes. In the extreme cases, we can exactly measure the initial phase space point and then choose its unique protocol. More generally, we assign a protocol to a region in the phase space, while the region can be arbitrary in principle. According to these observations, we infer that our main result may have potential applications in thermodynamics with the feedback control.

Actually, we can easily derive one of the Sagawa-Ueda equalities \cite{Sagawa_2010} based on our main result or its side products. We apply Eq.~(\ref{WPDFcal}) to a protocol $\lambda^y_t$ which corresponds to the measurement result $y$
\begin{equation}
P_\tau (W | y) = e^{\beta (W - \Delta F^y)} \int_{\mathfrak{S}_0} d\Gamma \bar P_\tau (-W,\Gamma | y) \frac{p_0 (\Gamma^\dagger | y)}{p^{eq}_0 (\Gamma^\dagger)} \; .
\label{WPDFcalforfc}
\end{equation}
Here the conditional initial distribution in the phase space $p_0 (\Gamma | y)$ satisfies
\begin{equation}
p_0 (\Gamma | y) p (y) = p (y | \Gamma) p^{eq}_0 (\Gamma) \; ,
\label{condprob}
\end{equation}
as long as the system is initially prepared in a thermal equilibrium state, with $p(y) \equiv \int_{\mathfrak{S}_0} d\Gamma p(y | \Gamma) p^{eq}_0 (\Gamma)$ to be the probability density that the measurement result turns out to be $y$. $p(y | \Gamma)$ is the conditional probability density that the measurement result is $y$ conditioned on the phase space point $\Gamma$. 
For simplicity, we assume that the measurement has the property $p(y^\dagger | \Gamma^\dagger ) = p(y | \Gamma)$ \cite{Sagawa_2010}. This relation is obviously satisfied for the error-free measurement $p(y | \Gamma) = \delta (y - \Gamma)$, and the Gaussian-error measurement $p(y | \Gamma) = (2 \pi \sigma^2)^{-\frac{1}{2}} e^{-|y - \Gamma|^2/2 \sigma^2}$, with $\Gamma = (\boldsymbol{r},\boldsymbol{p})$. Combining Eq.~(\ref{WPDFcalforfc}) and Eq.~(\ref{condprob}), we obtain
\begin{equation}
p(y) P_\tau (W | y) e^{- \beta (W - \Delta F^y)} = \int_{\mathfrak{S}_0} d\Gamma \bar P_\tau (-W,\Gamma | y) p (y | \Gamma^\dagger) \; .
\label{preSUE}
\end{equation}
After integrating Eq.~(\ref{preSUE}) over both $y$ and $W$, we get its integral version
\begin{equation}
\langle e^{-\beta (W - \Delta F)} \rangle = \eta \; ,
\label{SUE}
\end{equation}
where
\begin{equation}
\eta = \int_{\mathfrak{S}_0} dy d\Gamma p (y^\dagger | \Gamma)  \bar p_\tau (\Gamma | y) = \int_{\mathfrak{S}_0} dy \bar p (y^\dagger | y) \; .
\label{etadef}
\end{equation}
Here $\bar p_\tau (\Gamma | y) \equiv \int dW \bar P_\tau (W,\Gamma | y)$ is the final distribution of the phase space point for the backward process, and $\bar p (y' | y)$ is the probability density that the final phase space point is measured to be $y'$, after being driven by the protocol $\bar \lambda^y_t$ from the canonical time-reversed initial distribution.

The other main result in Ref.~\cite{Sagawa_2010} reads
\begin{equation}
\langle e^{-\beta (W - \Delta F) - I} \rangle = 1 \; ,
\label{SUE2}
\end{equation}
where $I$ is the initial mutual information between the system and the measurement device. However, in both Eq.~(\ref{SUE}) and Eq.~(\ref{SUE2}) it is required that the initial distribution must be a canonical distribution. To construct a more general information-involved FT, Sagawa and Ueda proposed
\begin{equation}
\langle e^{-\sigma + \Delta I} \rangle = 1
\label{SUE3}
\end{equation}
in Ref.~\cite{Sagawa_2012} as a generalization of the EPI. Here $\Delta I$ is the difference of the initial and the final mutual information; $\sigma \equiv \beta Q + \Delta s$ and $ \Delta s \equiv - \ln p_\tau (\Gamma_\tau) + \ln p_0 (\Gamma_0 )$, with $p_\tau (\Gamma) = \int dy p_\tau (\Gamma | y) p(y)$. The above two Sagawa-Ueda equalities (\ref{SUE2}) and (\ref{SUE3}) are actually contained in the unified IFT (\ref{uniIFT}), if we use the point of view in Ref.~\cite{Sagawa_2012} to regard the combination of the original system and the device as a composite system (see Appendix~\hyperref[app:uniIFTtoSUE]{F}). However, the counterpart of Eq.~(\ref{SUE}) as another generalization of the EPI seems to be unexplored so far. Now we can derive it quite straightforwardly in analogy to the derivation of Eq.~(\ref{SUE}). Following the same procedure as that from Eq.~(\ref{WPDFcalforfc}) to Eq.~(\ref{preSUE}), we obtain
\begin{equation}
p(y) P_\tau (\sigma | y) e^{-\sigma} = \int_{\mathfrak{S}_0} d\Gamma \bar P_\tau (-\sigma,\Gamma | y) p (y | \Gamma^\dagger) \; ,
\label{pregEPI}
\end{equation}
which indicates
\begin{equation}
\langle e^{-\sigma} \rangle = \eta \; .
\label{gEPI}
\end{equation}
The expression of $\eta$ is exactly the same as Eq.~(\ref{etadef}), but $\bar p (y' | y)$ must be associated with the initial distribution of the time-reversed process, which is the real time-reversal of the forward final distribution, i.e., $\bar p_0 (\Gamma) = p_\tau (\Gamma^\dagger)$, instead of a canonical one.

As an illustrative example of Eq.~(\ref{gEPI}), we consider the Szilard engine \cite{Szilard_1929}, as was used in Ref.~\cite{Sagawa_2010} to illustrate Eqs.~(\ref{SUE}) and (\ref{SUE2}). If the process is reversible and the measurement is error-free, we will always have $\Delta s = 0$ and $\sigma = \beta Q = - \ln 2$ (delta distribution), so that $\langle e^{-\sigma} \rangle = 2$. On the other hand, if $y = l$ ($r$), i.e., the particle is found in the left (right) half chamber, $\lambda^y_t$ will be the rightward (leftward) expansion. Thus $\bar \lambda^y_t$ will be the leftward (rightward) compression. It is easy to see that $\bar p (l | l)$ or $\bar p (r | r)$ ($l^\dagger = l$, $r^\dagger = r$, since $l$ or $r$ is related to the position that is invariant under time-reversal) is simply $1$, owing to the fact that the particle must be always in the left (right) half chamber after the leftward (rightward) compression. Hence, $\eta = \bar p (l | l) + \bar p (r | r) = 2$, and Eq.~(\ref{gEPI}) is indeed valid for such feedback control processes. What's more, even if the process is far from the quasistatic one, we will still have $\eta = 2$ due to the same analysis. This result generally implies $-\langle \sigma \rangle \leq \ln 2$, which is one aspect of the Landauer's principle \cite{Landauer_1961,Bennett_1982}.

\section{Conclusion}
\label{sec:conclusion}

In summary, we propose a refined unified FT (\ref{mainres}) under the DB condition. The refined unified FT is applicable to several thermodynamic variables, such as the heat, the work, the trajectory entropy production, and is even more refined than the DFT. 
Compared with the previous unified IFT and DFT \cite{Seifert_2008,Seifert_2012}, our refined unified FT (\ref{mainres}) eliminates the entanglement between the thermodynamic variable and the choice of the initial distributions in the phase space, thus is physically more natural and comprehensible. In particular, our refined unified FT generalizes the FTs of the work and the heat, such as the JE and the CFT, to arbitrary initial distributions (see Table \ref{table1}), as well as generates several new FTs that were previously not known to researchers in this field, for example, the Jarzynski equality to an arbitrary initial state (\ref{Wexpavg}). We also revisits the validity of the DFT of entropy production \cite{Esposito_2010,Thermalization,Van den Broeck_2015}. The price that we pay for the generalization to arbitrary initial distributions is that we need to know the joint distribution functions with the phase space point. Due to such kind of generalizations, our results might be valuable for the free energy recovering experiment and free energy computing as well as studying thermodynamics with information exchange, where the initial distributions are, to some extent, irregular. Based on the refined unified FT, we reproduce one (\ref{SUE}) of the Sagawa-Ueda equalities and derive a new generalized EPI (\ref{gEPI}) for the feedback control processes.

\acknowledgments
We thank Fei Liu for helpful discussions. H.T.Q. gratefully acknowledges support from the National Science Foundation of China under grant 11375012, and The Recruitment Program of Global Youth Experts of China.

\appendix

\section{Path Integral Derivation of the Main Result (\ref{mainres})}
\label{app:PIderivation}

In the path integral representation, the DB condition is reflected by the following relation
\begin{equation}
\beta Q[\Gamma_t] = \ln \frac{\mathcal{P}(\Gamma_t | \Gamma_0)}{\mathcal{\bar P}(\bar \Gamma_t | \bar \Gamma_0)} \; ,
\label{DBPI}
\end{equation}
which is a crucial building block in the later derivation of the main result. Here $\mathcal{P}(\Gamma_t | \Gamma_0)$ ($\mathcal{\bar P}(\bar \Gamma_t | \bar \Gamma_0)$) is the conditional probability density of the trajectory $\Gamma_t$ ($\bar\Gamma_t$) when the initial phase space point is known to be $\Gamma_0$ ($\bar\Gamma_0$) for the forward (backward) process. While Eq.~(\ref{DBPI}) has appeared in various references, such as its original discrete version in Ref.~\cite{Crooks_1998}, to make the paper self-contained, we briefly derive it from the basic trajectory definition of heat and the dynamic property (\ref{crucialdp}). To do this, we first consider a sufficiently short time interval $[t,t+dt]$ in the forward process, during which
\begin{equation}
p(\Gamma_{t+dt}|\Gamma_t)=\int_{\mathfrak{S}_{t+\frac{dt}{2}}}d\Gamma\delta(\Gamma-\Gamma_{t+dt})(1+ \mathcal{L}_{t+\frac{dt}{2}}dt)\delta (\Gamma-\Gamma_t) \; .
\label{Fdt}
\end{equation}
Here $p(\Gamma_{t_2}|\Gamma_{t_1})\equiv\int_{\mathfrak{S}_{t_2}}d\Gamma \delta(\Gamma-\Gamma_{t_2}) \mathcal{T}\{e^{\int^{t_2}_{t_1}dt \mathcal{L}_t}\}\delta(\Gamma-\Gamma_{t_1})$ is the conditional probability density that the phase space point at $t_2$ is $\Gamma_{t_2}$ starting from $\Gamma_{t_1}$ (a delta distribution) at $t_1$($<t_2$), and the terms of the order of magnitude of $(dt)^2$ are neglected. Similarly, for a short time interval $[\tau-t-dt,\tau-t]$ in the backward process, we have
\begin{equation}
\begin{split}
&\bar p(\bar \Gamma_{\tau-t}|\bar \Gamma_{\tau-t-dt})\\
=& \int_{\mathfrak{S}_{t+\frac{dt}{2}}} d\Gamma \delta (\Gamma-\bar \Gamma_{\tau-t}) (1+\mathcal{\bar L}_{\tau-t-\frac{dt}{2}}dt)\delta (\Gamma-\bar \Gamma_{\tau-t-dt})\\
=&\int_{\mathfrak{S}_{t+\frac{dt}{2}}}d\Gamma\delta(\Gamma-\Gamma_t)(1+\mathcal{L}^\dagger_{t+\frac{dt}{2}}dt)\delta (\Gamma-\Gamma_{t+dt})\; .
\end{split}
\label{Bdt}
\end{equation}
Combining Eqs.~(\ref{Fdt}) and (\ref{Bdt}) with the dynamic property (\ref{crucialdp}), we deduce that
\begin{equation}
\ln\frac{p(\Gamma_{t+dt}|\Gamma_t)}{\bar p(\bar \Gamma_{\tau-t}|\bar \Gamma_{\tau-t-dt})}
=-\beta\partial_\Gamma U_{t+\frac{dt}{2}}(\Gamma_{t+\frac{dt}{2}})\dot \Gamma_{t+\frac{dt}{2}} dt \; .
\end{equation}
Due to the Markovianity of the dynamics, the probability density of a trajectory can be expressed as $\mathcal{P}[\Gamma_t|\Gamma_0]=\lim_{K\to\infty}\prod^K_{k=1}p(\Gamma_{\frac{k}{K}\tau}| \Gamma_{\frac{k-1}{K}\tau})$. Therefore, we obtain
\begin{equation}
\begin{split}
\ln\frac{\mathcal{P}[\Gamma_t|\Gamma_0]}{\mathcal{\bar P}[\bar \Gamma_t|\bar \Gamma_0]}&=\lim_{K\to\infty}\sum^K_{k=1}
\ln\frac{p(\Gamma_{\frac{k}{K}\tau}|\Gamma_{\frac{k-1}{K}\tau})}{\bar p(\bar \Gamma_{\frac{K-k+1}{K}\tau}|\bar \Gamma_{\frac{K-k}{K}\tau})} \\
&=-\beta\int^\tau_0 dt \partial_\Gamma U(\Gamma_t)\dot\Gamma_t=\beta Q[\Gamma_t]
\end{split}
\end{equation}

Now we start the actual derivation of Eq.~(\ref{mainres}). Recalling the definition of $P_\tau (A,\Gamma)$, in the path integral representation we have
\begin{equation}
P_\tau (A,\Gamma) = \int \mathcal{D}[\Gamma_t] \mathcal{P}(\Gamma_t | \Gamma_0) p_0(\Gamma_0) \delta (A[\Gamma_t] - A) \delta (\Gamma_\tau - \Gamma) \; ,
\label{jPDFPIdef}
\end{equation}
with $\mathcal{D}[\Gamma_t] \equiv \lim_{K \to \infty} \prod^K_{k = 0} d\Gamma_{k\tau/K}$. In this manner, the denominator on the l. h. s. of Eq.~(\ref{mainres}), denoted by $L$ for convenience, can be rewritten as
\begin{equation}
\begin{split}
L = \int \mathcal{D}[\Gamma_t] \int_{\mathfrak{S}_\tau} d\Gamma e^{a_\tau (\Gamma)} \mathcal{P}(\Gamma_t | \Gamma_0) p_0(\Gamma_0) \bar p_0(\Gamma^\dagger)\\ \times \delta (A[\Gamma_t] - A) \delta (\Gamma_\tau - \Gamma) \; .
\end{split}
\end{equation}
By making use of Eq.~(\ref{DBPI}) and the property of the delta function, we obtain
\begin{equation}
\begin{split}
L = \int \mathcal{D}[\Gamma_t] e^{\beta Q[\Gamma_t] + a_\tau (\Gamma_\tau)} \mathcal{\bar P}(\bar \Gamma_t | \bar \Gamma_0) \bar p_0(\Gamma^\dagger_\tau) \\ \times p_0(\Gamma_0) \delta (A[\Gamma_t] - A) \; .
\end{split}
\end{equation}
Then we use the definition of $A$ (\ref{Adef}) to replace $\beta Q [\Gamma_t] + a_\tau (\Gamma_\tau)$ with $A[\Gamma_t] + a_0 (\Gamma_0)$, which leads to
\begin{equation}
\begin{split}
L = \int \mathcal{D}[\Gamma_t] e^{A[\Gamma_t] + a_0 (\Gamma_0)} \mathcal{\bar P}(\bar \Gamma_t | \bar \Gamma_0) \bar p_0(\Gamma^\dagger_\tau) \\ \times p_0(\Gamma_0) \delta (A[\Gamma_t] - A) \; .
\end{split}
\end{equation}
Again we use the property of the delta function, getting
\begin{equation}
\begin{split}
L = e^A \int \mathcal{D}[\Gamma_t] e^{a_0 (\Gamma_0)} \mathcal{\bar P}(\bar \Gamma_t | \bar \Gamma_0) \bar p_0(\Gamma^\dagger_\tau) \\  \times p_0(\Gamma_0) \delta (A[\Gamma_t] - A) \; .
\end{split}
\end{equation}
According to the definition of the time-reversal, it can be checked that  $\bar A[\bar \Gamma_t] = - A[\Gamma_t]$, e.g., $ Q[\bar \Gamma_t ] = - Q [\Gamma_t]$, so that
\begin{equation}
\begin{split}
L = e^A \int \mathcal{D}[\bar \Gamma_t] \int_{\mathfrak{S}_0} d\Gamma e^{\bar a_\tau (\bar \Gamma_\tau)} \mathcal{\bar P}(\bar \Gamma_t | \bar \Gamma_0) \bar p_0(\bar \Gamma_0) \\
\times p_0(\bar \Gamma^\dagger_\tau) \delta (\bar A[\bar \Gamma_t] + A) \delta (\bar \Gamma_\tau - \Gamma) \; .
\end{split}
\label{deltainsert}
\end{equation}
The insertion of the delta function aims at constructing the joint distribution function of $\bar A$ for the time-reversed process and the final phase space point. Particularly, we rewrite Eq.~(\ref{deltainsert}) as
\begin{equation}
\begin{split}
L = e^A \int \mathcal{D}[\bar \Gamma_t] \int_{\mathfrak{S}_0} d \Gamma e^{\bar a_\tau (\Gamma)} \mathcal{\bar P}(\bar \Gamma_t | \bar \Gamma_0) \bar p_0(\bar \Gamma_0) \\ \times p_0(\Gamma^\dagger) \delta (\bar A[\bar \Gamma_t] - (-A)) \delta (\bar \Gamma_\tau - \Gamma) \; .
\end{split}
\label{reconstruct}
\end{equation}
Recall the path integral representation of the joint distribution function (\ref{jPDFPIdef}), Eq.~(\ref{reconstruct}) actually leads to
\begin{equation}
L = e^A \int_{\mathfrak{S}_0} d\Gamma \bar P_\tau (-A,\Gamma) p_0(\Gamma^\dagger) e^{\bar a_\tau (\Gamma)} \; ,
\end{equation}
which is nothing but the numerator on the l. h. s. of Eq.~(\ref{mainres}) multiplied by $e^{A}$.

\section{Validity of the Dynamic Property (\ref{crucialdp}) for the Langevin Dynamics}
\label{app:cdpforLD}

We focus on the one-dimensional case here. The generalization to higher dimensions is straightforward, though anisotropic effect may emerge. For the overdamped Langevin dynamics, the generator $\mathcal{L}_t$ in the FPE (\ref{FPE}) reads \cite{Sekimoto_2010}
\begin{equation}
\mathcal{L}_t = \frac{1}{\gamma} \partial_x (\partial_x U_t + \beta^{-1} \partial_x) \; .
\end{equation}
Here $\gamma$ is the viscous friction coefficient; $U_t \equiv V_t (x)$ only depends on the position, indicating
\begin{equation}
\mathcal{L}^\dagger_t = \mathcal{L}_t \; .
\end{equation}
Notice that $\int^{+ \infty}_{- \infty} dx f(x) \frac{d}{dx} g(x) = - \int^{+ \infty}_{- \infty} dx g(x) \frac{d}{dx} f(x)$ for any normalizable functions $f(x)$ and $g(x)$ (so that $\lim_{x \to \pm \infty} f(x)g(x) = 0$), the transpose operator of $\mathcal{L}_t$ should be
\begin{equation}
\mathcal{L}^T_t = \frac{1}{\gamma} (- \partial_x U_t + \beta^{-1} \partial_x) \partial_x  \; .
\end{equation}
Before checking the dynamic property (\ref{crucialdp}), we introduce the following two useful relations
\begin{equation}
\begin{split}
e^{\beta U_t} \partial_x e^{- \beta U_t} =& \partial_x - \beta \partial_x U_t  \; .\\
e^{\beta U_t} \partial^2_x e^{- \beta U_t} =& (\partial_x - \beta \partial_x U_t)^2 \\
=& \partial^2_x - 2\beta \partial_x U_t \partial_x - \beta \partial^2_x U_t + \beta^2 (\partial_x U_t)^2\; .
\end{split}
\end{equation}
With these relations in hand, we start to calculate the l. h. s. of Eq.~(\ref{crucialdp})
\begin{equation}
\begin{split}
&e^{\beta U_t} \mathcal{L}^\dagger_t e^{- \beta U_t} \\
=& \frac{1}{\gamma} [\partial^2_x U_t + \partial_x U_t \partial_x - \beta (\partial_x U_t)^2 \\
&+ \beta^{-1} \partial^2_x - 2 \partial_x U_t \partial_x - \partial^2_x U_t + \beta (\partial_x U_t)^2] \\
= &\frac{1}{\gamma} (- \partial_x U_t \partial_x + \beta^{-1} \partial^2_x ) =  \mathcal{L}^T_t \; .
\end{split}
\end{equation}
Thus, Eq.~(\ref{crucialdp}) has been confirmed to be valid for the overdamped Langevin dynamics.

Let's move on to the underdamped Langevin dynamics. The generator $\mathcal{L}_t$ reads \cite{Sekimoto_2010}
\begin{equation}
\mathcal{L}_t = - \frac{p}{m} \partial_x + \partial_p (\partial_x U_t + \gamma \frac{p}{m}) + \gamma \beta^{-1} \partial^2_p \; ,
\end{equation}
based on which we can obtain its time-reversed operator and its transpose operator
\begin{equation}
\begin{split}
\mathcal{L}^\dagger_t =& \frac{p}{m} \partial_x + \partial_p (- \partial_x U_t + \gamma \frac{p}{m}) + \gamma \beta^{-1} \partial^2_p \; ,\\
\mathcal{L}^T_t =& \frac{p}{m} \partial_x - (\partial_x U_t + \gamma \frac{p}{m}) \partial_p + \gamma \beta^{-1} \partial^2_p \; .
\end{split}
\end{equation}
Here $U_t \equiv p^2/2m + V_t (x)$ depends on both the position and the momentum. Again we introduce two useful relations first
\begin{equation}
\begin{split}
e^{\beta U_t} \partial_p e^{- \beta U_t} =& \partial_p - \beta \frac{p}{m} \;\\
e^{\beta U_t} \partial^2_p e^{- \beta U_t} =& (\partial_p - \beta \frac{p}{m})^2 \\
=& \partial^2_p - 2\beta \frac{p}{m} \partial_p - \frac{\beta}{m} + (\frac{\beta p}{m})^2 \; .
\end{split}
\end{equation}
Then we calculate the l. h. s. of Eq.~(\ref{crucialdp})
\begin{equation}
\begin{split}
&e^{\beta U_t} \mathcal{L}^\dagger_t e^{- \beta U_t}\\
 = &\frac{p}{m} (\partial_x - \beta \partial_x U_t) \\
&+ \frac{\gamma}{m} + (- \partial_x U_t + \gamma \frac{p}{m})(\partial_p - \beta \frac{p}{m}) \\
&+ \gamma \beta^{-1} \partial^2_p - 2\gamma \frac{p}{m} \partial_p - \frac{\gamma}{m} + \gamma \beta (\frac{p}{m})^2 \\
=& \frac{p}{m} \partial_x - \partial_x U_t \partial_p - \gamma \frac{p}{m} \partial_p + \gamma \beta^{-1} \partial^2_p =  \mathcal{L}^T_t \; .
\end{split}
\end{equation}
Thus, Eq.~(\ref{crucialdp}) has also been confirmed to be valid for the underdamped Langevin dynamics.

\section{General equation of motion for the Joint Distribution Function}
\label{app:generalEOM}

Instead of the definition in the main text (\ref{Adef}), let¡¯s consider a thermodynamic variable associated with a process generally expressed as
\begin{equation}
A[\Gamma_t] = \int^{\tau}_0 dt \partial_t w_t (\Gamma_t) + \int^{\tau}_0 dt \dot \Gamma_t \partial_{\Gamma} q_t (\Gamma_t) \; .
\label{gAdef}
\end{equation}
Here $w_t (\Gamma)$ and $q_t (\Gamma)$ can be arbitrary time-dependent functions with respect to the phase space point. We will show that if the generator is $\mathcal{L}_t$, the equation of motion for $P_t (A,\Gamma)$ will be
\begin{equation}
\partial_{t} P_t (A,\Gamma) = [e^{-q_t(\Gamma) \partial_{A}} \mathcal{L}_t e^{q_t(\Gamma) \partial_{A}} - \partial_{t} w_t(\Gamma) \partial_{A}] P_t (A,\Gamma) \; .
\label{gEOM}
\end{equation}
It is instructive to first consider a simple case that $q_t (\Gamma) = 0$. In this case, $P_{t + dt} (A,\Gamma)$ is related to $P_t (A,\Gamma)$ via the following relation \cite{Sekimoto_2010} (terms with the magnitude of $(dt)^2$ are ignored)
\begin{equation}
\begin{split}
P_{t + dt} (A,\Gamma) =& (1 + \mathcal{L}_t dt) P_t (A - \partial_{t} w_t(\Gamma) dt,\Gamma) \\
=& \mathcal{L}_t P_t (A,\Gamma) dt + P_t (A - \partial_{t} w_t(\Gamma) dt,\Gamma) \; ,
\label{zeroq}
\end{split}
\end{equation}
which implies
\begin{equation}
\partial_{t} P_t (A,\Gamma) = [\mathcal{L}_t - \partial_{t} w_t(\Gamma) \partial_{A}] P_t (A,\Gamma) \; .
\label{preFKE}
\end{equation}
This result is familiar to us since it is the precursor of the Feynman-Kac formula before performing the integral transformation. In order to generalize to the case with $q_t (\Gamma) \neq 0$, we rewrite the first term in the rightmost of Eq.~(\ref{zeroq}) as
\begin{equation}
\begin{split}
\mathcal{L}_t P_t (A,\Gamma) dt = &\int_{\mathfrak{S}_t} d\Gamma' d\Gamma'' P_t (A,\Gamma'') \\
& \times \delta(\Gamma'-\Gamma) \mathcal{L}'_t \delta(\Gamma'-\Gamma'') dt \; ,
\end{split}
\end{equation}
where $\mathcal{L}'_t$ acts only on $\Gamma'$. Now the physical meaning is transparent: this is the contribution to $P_{t+dt} (A,\Gamma)$ due to motions in the phase space, consisting of both parts that come from other phase space points ($\Gamma'' \neq \Gamma$) and leave $\Gamma$ ($\Gamma'' = \Gamma$, this is necessary to correct the second term in the rightmost of Eq.~(\ref{zeroq}), the contribution due to the temporal variation of $w_t(\Gamma)$). Based on such an interpretation, when a nonzero $q_t (\Gamma)$ appears, we simply modify Eq.~(\ref{zeroq}) as
\begin{equation}
\begin{split}
&P_{t + dt} (A,\Gamma) = \int_{\mathfrak{S}_t} d\Gamma' d\Gamma'' P_t (A - q_t (\Gamma) + q_t (\Gamma''),\Gamma'') \\
&\times \delta(\Gamma'-\Gamma) \mathcal{L}'_t \delta(\Gamma'-\Gamma'') dt + P_t (A - \partial_{t} w_t(\Gamma) dt,\Gamma)  \; ,
\end{split}
\label{nonzeroq}
\end{equation}
owing to the second term in Eq.~(\ref{gAdef}). Since the first term in Eq.~(\ref{nonzeroq}) is already of the order of the magnitude of $dt$, we don't have to further add modifications like $\partial_t w_t (\Gamma)dt$ or $\partial_t q_t (\Gamma)dt$ in addition to $- q_t (\Gamma) + q_t (\Gamma'' )$, which will merely result in differences of the order of the magnitude of $(dt)^2$. By using the identity $e^{a \frac{d}{dx}} f(x) = f(x + a)$ as well as the property of the delta function, we can simplify Eq.~(\ref{nonzeroq}) as
\begin{equation}
\begin{split}
P_{t + dt} (A,\Gamma) = e^{- q_t (\Gamma) \partial_A} \mathcal{L}_t e^{q_t (\Gamma) \partial_A} P_t (A,\Gamma) dt \\
+ P_t (A - \partial_{t} w_t(\Gamma) dt,\Gamma) \; ,
\end{split}
\end{equation}
which finally leads to Eq.~(\ref{gEOM}).

Now let's return to the functional in the main text (\ref{Adef}). By using the identity
\begin{equation}
a_\tau (\Gamma_\tau) - a_0 (\Gamma_0) = \int^\tau_0 dt \partial_t a_t (\Gamma_t) + \int^\tau_0 dt \dot \Gamma_t \partial_\Gamma a_t (\Gamma_t) \; ,
\end{equation}
we find that Eq.~(\ref{Adef}) can be rewritten as
\begin{equation}
A[\Gamma_t] = \int^{\tau}_0 dt \partial_t a_t (\Gamma_t) + \int^{\tau}_0 dt \dot \Gamma_t \partial_{\Gamma} [a_t (\Gamma_t) - \beta U_t (\Gamma_t)] \; .
\label{c9}
\end{equation}
Comparing Eq.~(\ref{c9}) with Eq.~(\ref{gAdef}), we have $w_t (\Gamma) = a_t (\Gamma)$ and $q_t (\Gamma) = a_t (\Gamma) - \beta U_t (\Gamma)$. Substituting them into Eq.~(\ref{gEOM}), we get the equation of motion in the main text (\ref{EOM}).

\section{FT of work (\ref{dworkFT}) for Driven Isolated Quantum Systems}
\label{app:qworkFT}

We generally denote the Hamiltonian of an isolated quantum system by $\mathcal{H}_t$, whose time-reversal is determined by $\mathcal{\bar H}_t \equiv \Theta \mathcal{H}_{\tau - t} \Theta^{-1}$, with $\Theta$ to be the antiunitary time-reversal operator. To be specific, the quantum version of Eq.~(\ref{dworkFT}) can be written as
\begin{equation}
\begin{split}
&\sum_n P_\tau (W, | n \rangle) \bar p_0 (\Theta | n \rangle) e^{\beta E^n_\tau} \\
= &e^{\beta W} \sum_{\bar m} \bar P_\tau (-W, | \bar m \rangle) p_0 (\Theta^{-1} | \bar m \rangle) e^{\beta \bar E^{\bar m}_\tau} \; .
\end{split}
\label{qworkFT}
\end{equation}
Here $| n \rangle$ ($| \bar m \rangle$) is an eigenstate of $\mathcal{H}_\tau$ ($\mathcal{\bar H}_\tau$) with the eigenenergy $E^n_\tau$ ($\bar E^{\bar m}_\tau$); $p_0 (\Theta^{-1} | \bar m \rangle)$ ($\bar p_0 (\Theta | n \rangle)$) is the probability that the initial state for the forward (backward) process is measured to be $\Theta^{-1}| \bar m \rangle$ ($\Theta | n \rangle$), which is obviously an eigenstate of $\mathcal{H}_0$ ($\mathcal{\bar H}_0$) due to the former definitions. Such probability can be evaluated in terms of the initial density operator $\varrho_0$ ($\bar \varrho_0$) via
\begin{equation}
\begin{split}
p_0 (\Theta^{-1} | \bar m \rangle) =& \langle \bar m | \Theta \varrho_0 \Theta^{-1} | \bar m \rangle \; ,\\
\bar p_0 (\Theta | n \rangle) =& \langle n | \Theta^{-1} \bar \varrho_0 \Theta | n \rangle \; ,
\end{split}
\end{equation}
where $[\varrho_0 , \mathcal{H}_0] = [\bar \varrho_0 , \mathcal{\bar H}_0 ] = 0$ due to its structure assumed in the main text. Accordingly, $[\Theta \varrho_0 \Theta^{-1}, \mathcal{\bar H}_\tau] = [\Theta^{-1} \bar \varrho_0 \Theta, \mathcal{H}_\tau] = 0$ holds subsequently. Based on the two-point projection measurement definition of quantum work, the joint distribution functions should be
\begin{equation}
\begin{split}
P_\tau (W, | n \rangle) =& \sum_m |\langle n | \mathcal{U}_{\tau,0} | m \rangle|^2 p_0 (| m \rangle) \delta(W - E^n_\tau + E^m_0) \; ,\\
\bar P_\tau (W, | \bar m \rangle) =& \sum_{\bar n} |\langle \bar m | \mathcal{\bar U}_{\tau,0} | \bar n \rangle|^2 \bar p_0 (| \bar n \rangle) \delta(W - E^{\bar m}_\tau + E^{\bar n}_0) \; ,
\end{split}
\end{equation}
where $p_0 (| m \rangle)$ ($\bar p_0 (| \bar n \rangle)$) can also be related to $\varrho_0$ ($\bar \varrho_0$) by $\langle m | \varrho_0 | m \rangle$ ($\langle \bar n | \bar \varrho_0 | \bar n \rangle$).

To prove Eq.~(\ref{qworkFT}), we again take the characteristic function-based approach, which has been widely used in the studies of quantum thermodynamics \cite{Talkner_2007,Talkner_2011}. We take the inverse Fourier transformation on both sides of Eq.~(\ref{qworkFT}), and obtain
\begin{equation}
\begin{split}
&\int^{+ \infty}_{- \infty} dW e^{i \mu W} \sum_n P_\tau (W, | n \rangle) \bar p_0 (\Theta | n \rangle) e^{\beta E^n_\tau} \\
=& \sum_n \sum_m |\langle n | \mathcal{U}_{\tau,0} | m \rangle|^2 p_0 (| m \rangle) \bar p_0 (\Theta | n \rangle) e^{(i \mu + \beta) E^n_\tau - i \mu E^m_0} \\
=& \sum_n \sum_m \langle n | e^{i \nu \mathcal{H}_\tau} \mathcal{U}_{\tau,0} \varrho_0 | m \rangle \langle m | e^{i \mu \mathcal{H}_0} \mathcal{U}_{0,\tau} \Theta^{-1} \bar \varrho_0 \Theta | n \rangle \\
=& \mathrm{Tr} [e^{i \nu \mathcal{H}_\tau} \mathcal{U}_{\tau,0} \varrho_0 e^{- i \mu \mathcal{H}_0} \mathcal{U}_{0,\tau} \Theta^{-1} \bar \varrho_0 \Theta] \; , \\
&\int^{+ \infty}_{- \infty} dW e^{\beta W} e^{i \mu W} \sum_{\bar m} \bar P_\tau (-W, | \bar m \rangle) \bar p_0 (\Theta^{-1} | \bar m \rangle) e^{\beta \bar E^{\bar m}_\tau} \\
=& \sum_{\bar m} \sum_{\bar n} |\langle \bar m | \mathcal{\bar U}_{\tau,0} | \bar n \rangle|^2 \bar p_0 (| \bar n \rangle) p_0 (\Theta^{-1} | \bar m \rangle) e^{(i \mu + \beta) \bar E^{\bar n}_0 - i \mu \bar E^{\bar m}_\tau} \\
= &\sum_{\bar m} \sum_{\bar n} \langle \bar m | e^{- i \mu \mathcal{\bar H}_\tau} \mathcal{\bar U}_{\tau,0} \bar \varrho_0 | \bar n \rangle \langle \bar n | e^{i \nu \mathcal{\bar H}_0} \mathcal{\bar U}_{0,\tau} \Theta \varrho_0 \Theta^{-1} | \bar m \rangle \\
= &\mathrm{Tr} [e^{- i \mu \mathcal{\bar H}_\tau} \mathcal{\bar U}_{\tau,0} \bar \varrho_0 e^{i \nu \mathcal{\bar H}_0} \mathcal{\bar U}_{0,\tau} \Theta \varrho_0 \Theta^{-1}] \; .
\end{split}
\end{equation}
Here $\mathcal{U}_{t,t'}$ ($\mathcal{\bar U}_{t,t'}$) is the time-evolution operator for the forward (backward) process, governed by the Schr\"{o}dinger equation $i \hbar \partial_t \mathcal{U}_{t,t'} = \mathcal{H}_t \mathcal{U}_{t,t'}$, $\mathcal{U}_{t',t'} \equiv \mathcal{I}$ ($i \hbar \partial_t \mathcal{\bar U}_{t,t'} = \mathcal{\bar H}_t \mathcal{\bar U}_{t,t'}$, $\mathcal{\bar U}_{t',t'} \equiv \mathcal{I}$), $\mathcal{I}$ is the identity operator; both $\mu$ and $\nu \equiv \mu - i \beta$ are generally complex numbers, i.e., $\mu$ and $\nu$ are unnecessarily their complex conjugates $\mu^\ast$, $\nu^\ast$, and are also independent to each other, owing to the arbitrariness of $\beta$. By making use of the algebraic properties of $\Theta$ and the trace (see Ref.~\cite{Talkner_2011} for details), especially $\mathcal{\bar U}_{t,t'} = \Theta \mathcal{U}_{\tau - t, \tau - t'} \Theta^{-1}$, $\Theta e^{- i \kappa \mathcal{H}_t} \Theta^{-1} = e^{ i \kappa^\ast \mathcal{\bar H}_{\tau - t}}$, $\mathrm{Tr} [\Theta^{-1} \mathcal{A} \Theta] = \mathrm{Tr} [\mathcal{A}^\dagger]$ and $\mathrm{Tr} [\mathcal{A} \mathcal{B}] = \mathrm{Tr} [\mathcal{B} \mathcal{A}]$ (be careful that this may be invalid if $\mathcal{A} = \Theta$), as well as the commutation relations $[\bar \varrho_0 , \mathcal{\bar H}_0 ] = 0$ and $[\Theta \varrho_0 \Theta^{-1}, \mathcal{\bar H}_\tau] = 0$, we have
\begin{equation}
\begin{split}
&\mathrm{Tr} [e^{i \nu \mathcal{H}_\tau} \mathcal{U}_{\tau,0} \varrho_0 e^{- i \mu \mathcal{H}_0} \mathcal{U}_{0,\tau} \Theta^{-1} \bar \varrho_0 \Theta] \\
=& \mathrm{Tr} [e^{i \nu \mathcal{H}_\tau} \Theta^{-1} \mathcal{\bar U}_{0,\tau} \Theta \varrho_0 e^{- i \mu \mathcal{H}_0} \Theta^{-1} \mathcal{\bar U}_{\tau,0} \bar \varrho_0 \Theta] \\
=& \mathrm{Tr} [\Theta^{-1} e^{-i \nu^\ast \mathcal{\bar H}_0} \mathcal{\bar U}_{0,\tau} \Theta \varrho_0 \Theta^{-1} e^{i \mu^\ast \mathcal{\bar H}_\tau} \mathcal{\bar U}_{\tau,0} \bar \varrho_0 \Theta] \\
=& \mathrm{Tr} [\bar \varrho_0 \mathcal{\bar U}_{0,\tau} e^{- i \mu \mathcal{\bar H}_\tau} \Theta \varrho_0 \Theta^{-1} \mathcal{\bar U}_{\tau,0} e^{i \nu \mathcal{\bar H}_0} ] \\
=& \mathrm{Tr} [e^{i \nu \mathcal{\bar H}_0} \mathcal{\bar U}_{0,\tau} \Theta \varrho_0 \Theta^{-1} e^{- i \mu \mathcal{\bar H}_\tau} \mathcal{\bar U}_{\tau,0} \bar \varrho_0] \\
=& \mathrm{Tr} [e^{- i \mu \mathcal{\bar H}_\tau} \mathcal{\bar U}_{\tau,0} \bar \varrho_0 e^{i \nu \mathcal{\bar H}_0} \mathcal{\bar U}_{0,\tau} \Theta \varrho_0 \Theta^{-1}] \; .
\end{split}
\label{qworkFTCF}
\end{equation}
Thus the inverse Fourier transformation of the two sides of Eq.~(\ref{qworkFT}) turns out to be the same. So far, the discrete version of the refined FT of work (\ref{dworkFT}) has been confirmed to be generally valid for driven isolated quantum systems.

As an example, if $\varrho_0 = e^{- \beta \mathcal{H}_0} / Z_0 (\beta)$ and $\bar \varrho_0 = e^{- \beta \mathcal{\bar H}_0} / \bar Z_0 (\beta)$, $Z_0 (\beta) \equiv \mathrm{Tr} [e^{- \beta \mathcal{H}_0}]$, $\bar Z_0 (\beta) \equiv \mathrm{Tr} [e^{- \beta \mathcal{\bar H}_0 }] = \mathrm{Tr} [e^{- \beta \mathcal{H}_\tau}] = Z_\tau (\beta)$ (due to $\mathrm{Tr} [\Theta^{-1} \mathcal{A} \Theta] = \mathrm{Tr} [\mathcal{A}^\dagger]$ and $\mathcal{H}^\dagger_\tau = \mathcal{H}_\tau$. This is the quantum analogy of footnote \cite{deltaF}), Eq.~(\ref{qworkFTCF}) will become $G(\mu)/ Z_\tau(\beta) = \bar G (- \mu + i \beta) / Z_0 (\beta)$. This means that the quantum CFT holds even for systems without the time-reversal symmetry, e.g., a charged particle subjected to a time-dependent magnetic field. This is a generalization of Ref.~\cite{Talkner_2011}, where $[\Theta , \mathcal{H}_t] = 0$ is assumed. However, we should be careful that the time-reversed Hamiltonian must be $\Theta \mathcal{H}_{\tau - t} \Theta^{-1}$, but usually not $\mathcal{H}_{\tau - t}$.

\section{Detailed Calculations on the breathing Brownian oscillator}
\label{app:Brownianoscillator}

To be consistent with Ref.~\cite{Speck_2011}, we use the generating function $\rho_t(\lambda,x)\equiv G_t(i\lambda,x)$ instead of the characteristic function $G_t(\mu,x)$. We first present the main result in Ref.~\cite{Speck_2011}, which focused on an arbitrary process driven by the protocol $k_t$ starting from the equilibrium state $p^{eq}_0(x)\equiv(k_0\beta/2\pi)^{\frac{1}{2}}e^{-\beta k_0x^2/2}$. By making the Gaussian ansatz
\begin{equation}
\rho_t(\lambda,x)=\sqrt{\frac{[\psi_\lambda(t)]^3}{2\pi\phi_\lambda(t)}} e^{-\frac{x^2\psi_\lambda(t)}{2\phi_\lambda(t)}} \; ,
\end{equation}
the Feynman-Kac formula can be self-consistently reduced to the following two first-order ordinary differential equations
\begin{equation}
\dot \psi_\lambda(t)=-\frac{\lambda \dot k_t}{2}\phi_\lambda(t)\;,\\
\label{ODE1}
\end{equation}
\begin{equation}
\dot \phi_\lambda(t)=-\frac{2k_t}{\gamma}\phi_\lambda(t)+\frac{2}{\beta\gamma}\psi_\lambda(t)-\frac{3\lambda\dot k_t}{2}\frac{\phi^2_\lambda(t)}{\psi_\lambda(t)}\;,
\label{ODE2}
\end{equation}
with the initial condition to be $\psi_\lambda(0)=1$ and $\phi_\lambda(0)=(\beta k_0)^{-1}$. To perform perturbative analysis, it is convenient to define $g_\lambda(t)\equiv\beta k_t \phi_\lambda(t)/\psi_\lambda(t)$. Based on Eqs.~(\ref{ODE1}) and (\ref{ODE2}), it can be checked that $g_\lambda(t)$ satisfies the following Raccati equation
\begin{equation}
\dot g_\lambda(t)=-\frac{2k_t}{\gamma}[g_\lambda(t)-1]+\frac{\dot k_t}{k_t}g_\lambda(t)[1-\frac{\lambda}{\beta}g_\lambda(t)] \; ,
\label{Riccati}
\end{equation}
with the initial condition to be $g_\lambda(0)=1$. Once $g_\lambda(t)$ is determined, $\psi_\lambda(\tau)=\langle e^{-\lambda W}\rangle$ can be obtained via $\ln\psi_\lambda(\tau)=-(\lambda/\beta)\int^\tau_0 dt g_\lambda(t) \dot k_t/2 k_t$. If we regard $\dot k_t$ in Eq.~(\ref{Riccati}) as a small quantity, the zeroth-order solution will simply read $g^{(0)}_\lambda(t)=1$, while the first-order correction should be $g^{(1)}_\lambda(t) = \gamma (1-\lambda /\beta) \dot k_t/2k^2_t$. Accordingly,$\ln\psi_\lambda(\tau)$ can be expressed as follows up to the first-order accuracy
\begin{equation}
\ln \psi_\lambda(\tau)=-\frac{\lambda}{2\beta}\ln\frac{k_\tau}{k_0} -\frac{\gamma\lambda}{4\beta} (1-\frac{\lambda}{\beta})\int^\tau_0 dt\frac{\dot k^2_t}{k^3_t}+O(\dot k^2_t) \; .
\end{equation}
In particular, for the protocol $k_t=k_0+\kappa t$, we have
\begin{equation}
\ln \psi^{(1)}_\lambda(\tau)=-\frac{\lambda}{2\beta}\ln\frac{k_\tau}{k_0} - \frac{\lambda}{8\beta} (1-\frac{\lambda}{\beta})(\frac{1}{k^2_0}-\frac{1}{k^2_\tau})\gamma\kappa \; .
\label{firstorder}
\end{equation}

Now let's start to calculate the generating function of the work distribution for the nonequilibrium initial state $p_0(x)=(k_0\beta'/2\pi)^{\frac{1}{2}}e^{-\frac{1}{2}\beta' k_0x^2}$. Using Eq.~(\ref{WGFcal}), we obtain
\begin{equation}
\begin{split}
&\langle e^{-\lambda W}\rangle_{p_0(x)} \\
=& e^{-\beta \Delta F}\int^{+\infty}_{-\infty}dx \bar\rho_\tau(\beta-\lambda,x) \frac{p_0(x)}{p^{eq}_0(x)}\\
=&\sqrt{\frac{\beta' k_0[\bar\psi_{\beta-\lambda}(\tau)]^3}{2\pi\beta k_\tau\bar\phi_{\beta-\lambda}(\tau)}}\int^{+\infty}_{-\infty}dx
e^{-\frac{x^2}{2}[\frac{\bar\psi_{\beta-\lambda}(\tau)}{\bar\phi_{\beta-\lambda}(\tau)}+k_0(\beta'-\beta)]}\\
=&\sqrt{\frac{k_0}{k_\tau}}[\frac{\beta}{\beta'}+(1-\frac{\beta}{\beta'})\bar g_{\beta-\lambda}(\tau)]^{-\frac{1}{2}}\bar \psi_{\beta-\lambda}(\tau) \; ,
\label{GeneratingF}
\end{split}
\end{equation}
where all the quantities with an overline must be associated with the time-reversed protocol $\bar k_t\equiv k_{\tau -t}$. After the first-order approximation, Eq.~(\ref{GeneratingF}) becomes
\begin{equation}
\begin{split}
&\ln \langle e^{-\lambda W}\rangle_{p_0(x)}\\
 = &\frac{1}{2}\ln\frac{k_0}{k_\tau}-\frac{1}{2}(1-\frac{\beta}{\beta'})\bar g^{(1)}_{\beta-\lambda}(\tau)+\ln \bar \psi^{(1)}_{\beta-\lambda}(\tau)+O(\kappa^2)\\
=&\ln \psi^{(1)}_\lambda(\tau)-\frac{1}{2}(1-\frac{\beta}{\beta'})\bar g^{(1)}_{\beta-\lambda}(\tau)+O(\kappa^2) \; ,
\end{split}
\end{equation}
where $\bar g^{(1)}_{\beta-\lambda}(\tau)=-\gamma\kappa\lambda/(2k^2_0\beta)$, and the fluctuation-dissipation relation $\ln \psi^{(1)}_\lambda(\tau)=-\beta\Delta F +\ln \bar\psi^{(1)}_{\beta - \lambda}(\tau)$ is used. Since the generating function in the form $\ln\langle e^{-\lambda W}\rangle=-\lambda\langle W\rangle+\lambda^2\sigma^2_W/2$ must correspond to a Gaussian distribution centred at $\langle W\rangle$ and with variance $\sigma^2_W$, we finally obtain the results (\ref{meanwork}) in the main text.

\section{Derivation of the Sagawa-Ueda equalities (\ref{SUE2}) and (\ref{SUE3}) from the Unified IFT}
\label{app:uniIFTtoSUE}

We consider the extended phase space as the direct (Cartesian) product of the phase spaces of the system and the measurement device. The phase space point in such extended space can be denoted by $\Sigma_t \equiv (\Gamma_t,y)$, with $\Gamma_t$ and $y$ to be the components of the system and the device respectively \cite{measurement}. The unified IFT of the composite system reads
\begin{equation}
\langle \frac{p^a_\tau (\Sigma_\tau)}{p_0 (\Sigma_0)}e^{- \Delta s_m} \rangle = 1 \; .
\label{compositeEPI}
\end{equation}
This is the key relation that we will use.

To derive Eq.~(\ref{SUE2}), we choose
\begin{equation}
\begin{split}
p_0 (\Sigma_0) &= p (y | \Gamma_0 ) \frac{e^{- \beta U_0 (\Gamma_0)}}{Z_0 (\beta)} \; ,\\
p^a_\tau (\Sigma_\tau) &= p (y) \frac{e^{- \beta U^y_\tau (\Gamma_\tau)}}{Z^y_\tau (\beta)} \; .
\end{split}
\label{terminaldistribution}
\end{equation}
By substituting Eq.~(\ref{terminaldistribution}) into Eq.~(\ref{compositeEPI}), we obtain
\begin{equation}
\langle \frac{Z_0 (\beta)}{Z^y_\tau (\beta)} e^{- \beta [Q + U^y_\tau (\Gamma_\tau) - U_0 (\Gamma_0)] - \ln [ p (y | \Gamma_0 ) / p (y)]} \rangle = 1 \; .
\label{preSUE2}
\end{equation}
Here the DB condition, and thus $\Delta s_m = \beta Q$ has been assumed. To simplify Eq.~(\ref{preSUE2}), we make use of the first law $W[\Gamma_t] = Q[\Gamma_t] + U_\tau (\Gamma_\tau) - U_0 (\Gamma_0)$ as well as the expression of the free energy difference $\Delta F^y \equiv - \beta^{-1} \ln [Z_\tau^y (\beta) / Z_0 (\beta)]$, and further define the state function of the mutual information $I(\Sigma_0) \equiv \ln [p(y | \Gamma_0) / p(y)]$. Combining these relations we obtain
\begin{equation}
\langle e^{-\beta (W - \Delta F) - I} \rangle = 1 \; .
\end{equation}

To derive Eq.~(\ref{SUE3}), we choose $p^a_\tau (\Sigma_\tau)$ to be the real final distribution $p_\tau (\Sigma_\tau)$. So we get
\begin{equation}
\langle e^{-\Delta s_{tot}} \rangle = 1 \; ,
\end{equation}
which is nothing but the EPI for the composite system. Concretely, we have
\begin{equation}
\Delta s_{tot} = \beta Q - \ln p_\tau (\Sigma_\tau) + \ln p_0 (\Sigma_0) \; ,
\end{equation}
as long as the DB condition holds. To distinguish two contributions to $\Delta s_{tot}$: (i) the correlation between the system and the device, and (ii) the entropy production of the system and in the medium, we define the mutual information at the initial and the final stages
\begin{equation}
I_i \equiv \ln \frac{p_0 (\Sigma_0)}{p_0 (\Gamma_0) p(y)} \; ,
I_f \equiv \ln \frac{p_\tau (\Sigma_\tau)}{p_\tau (\Gamma_\tau) p(y)} \; .
\end{equation}
Then we can obtain another decomposition identity of $\Delta s_{tot}$ as follows
\begin{equation}
\Delta s_{tot} = \beta Q - \ln p_\tau (\Gamma_\tau) + \ln p_0 (\Gamma_0) - I_f + I_i = \sigma - \Delta I \; ,
\end{equation}
which leads to
\begin{equation}
\langle e^{-\sigma + \Delta I} \rangle = 1 \; .
\end{equation}

It should be mentioned that $p_0(\Sigma)$ and $p^a_\tau(\Sigma)$ are tacitly assumed to be nonzero for any $\Sigma$ to guarantee the validity of the two Sagawa-Ueda equalities in this appendix. However, this assumption is usually not satisfied for error-free measurements or a rigorously localized initial state, thus the two Sagawa-Ueda equalities may break down in these cases \cite{Ueda_2014}. Recent researches have shown that by adding an extra modification term on the exponential, the Sagawa-Ueda equalities can be generalized to be applicable to the feedback control processes with error-free measurements \cite{Ueda_2014b}.


\begin{thebibliography}{99}
\bibitem{TRdef} To see why we choose such a definition instead of $\bar a_t (\Gamma) \equiv a_{\tau - t} (\Gamma)$, we simply consider an example that $a_t (\Gamma) = p_t (\Gamma)$ is the distribution function of the phase space point. If we want to obtain $\bar p_0 (\Gamma) = p_\tau (\Gamma^\dagger)$, the standard preparation of the initial distribution for the time-reversed process, we must follow the definition in the main text. Such a definition may also be heuristic for the potential generalization to the cases with nonconservative force, such as Lorentz force, which will break the symmetry $U_t (\Gamma) = U_t (\Gamma^\dagger )$.
\bibitem{acphsp} It is necessary to emphasize that the accessible phase space $\mathfrak{S_t}$ may change with time $t$, e.g., a molecule confined in an expanding piston. However, at any given time $t$, the accessible phase space of $\Gamma_t$ is always assumed to be the same as that of $\Gamma^\dagger_t$, namely $\Gamma_t\in\mathfrak{S_t}$ if and only if $\Gamma^\dagger_t\in\mathfrak{S_t}$.
\bibitem{deltaF} Remember that we have assumed that $\Gamma\in\mathfrak{S}$ if and only if $\Gamma^\dagger\in\mathfrak{S}$. Since the time-reversal operation of the phase space point does not change the size of volume element, we have $\int_{\mathfrak{S}} d\Gamma f(\Gamma) = \int_{\mathfrak{S}} d\Gamma f(\Gamma^\dagger)$ for any normalizable function $f(\Gamma)$ defined in the phase space $\mathfrak{S}$. In particular, $Z_\tau (\beta) = \bar Z_0 (\beta)$, where $\mathfrak{S}=\mathfrak{S}_\tau$, $f(\Gamma) = e^{-\beta U_\tau (\Gamma)}$ and $f(\Gamma^\dagger) = e^{-\beta U_\tau (\Gamma^\dagger)} = e^{-\beta \bar U_0 (\Gamma)}$. This is true even if $U_t (\Gamma) \neq U_t (\Gamma^\dagger)$, though we assume $U_t (\Gamma) = U_t (\Gamma^\dagger)$ througout the paper.
\end{thebibliography}
\end{document}